\documentclass[twocolumn,tighten]{aastex631}
\usepackage{graphicx}
\usepackage{amsmath}
\usepackage{natbib}
\usepackage{amssymb}

\usepackage{sidecap}

\shorttitle{Angular Momentum Transport in Rotating PNS and the Fate of BNS Merger Remnants}
\shortauthors{B.~Margalit, A.~S.~Jermyn, B.~D.~Metzger, L.~R.~Roberts, E.~Quataert}

\begin{document}

\newcommand{\be}{\begin{equation}}
\newcommand{\ee}{\end{equation}}

\newcommand  \gcc {~\mathrm{g}~\mathrm{cm}^{-3}}
\newcommand  \E {\times 10^}

\title{Angular Momentum Transport in Proto-Neutron Stars and the Fate of Neutron Star Merger Remnants}

\correspondingauthor{Ben Margalit}
\email{benmargalit@berkeley.edu}

\author[0000-0001-8405-2649]{Ben Margalit}
\altaffiliation{NASA Einstein Fellow}
\affiliation{Astronomy Department and Theoretical Astrophysics Center, University of California, Berkeley, Berkeley, CA 94720, USA}

\author[0000-0001-5048-9973]{Adam S.~Jermyn}
\affil{Center for Computational Astrophysics, Flatiron Institute, 162 5th Ave, New York, NY 10010, USA} 

\author[0000-0002-4670-7509]{Brian D.~Metzger}
\affil{Department of Physics and Columbia Astrophysics Laboratory, Columbia University, New York, NY 10027, USA}
\affil{Center for Computational Astrophysics, Flatiron Institute, 162 5th Ave, New York, NY 10010, USA} 

\author[0000-0001-7364-7946]{Luke F.~Roberts}
\affil{Computer, Computational, and Statistical Sciences Division, Los Alamos National Laboratory, Los Alamos, NM, 87545, USA} 

\author[0000-0001-9185-5044]{Eliot Quataert}
\affil{Department of Astrophysical Sciences, Princeton University, Princeton, NJ 08544, USA} 

\begin{abstract} 
Both the core collapse of rotating massive stars, and the coalescence of neutron star (NS) binaries, result in the formation of a hot, differentially rotating NS remnant. The timescales over which differential rotation is removed by internal angular-momentum transport processes (`viscosity') has key implications for the remnant's long-term stability and the NS equation-of-state (EOS). Guided by a non-rotating model of a cooling proto-NS, we estimate the dominant sources of viscosity using an externally imposed angular velocity profile $\Omega(r)$. Although the magnetorotational instability provides the dominant source of effective viscosity at large radii, convection and/or the Spruit-Tayler dynamo dominate in the core of merger remnants where $d\Omega/dr \geq 0$. Furthermore, the viscous timescale in the remnant core is sufficiently short that solid body rotation will be enforced faster than matter is accreted from rotationally-supported outer layers. Guided by these results, we develop a toy model for how the merger remnant core grows in mass and angular momentum due to accretion. We find that merger remnants with sufficiently massive and slowly rotating initial cores may collapse to black holes via envelope accretion, even when the total remnant mass is less than the usually considered threshold $\approx 1.2 M_{\rm TOV}$ for forming a stable solid-body rotating NS remnant (where $M_{\rm TOV}$ is the maximum non-rotating NS mass supported by the EOS). This qualitatively new picture of the post-merger remnant evolution and stability criterion has important implications for the expected electromagnetic counterparts from binary NS mergers and for multi-messenger constraints on the NS EOS.
\end{abstract}

\keywords{
Neutron stars (1108); 
Core-collapse supernovae (304); 
Gravitational waves (678); 
Stellar physics (1621); 
High energy astrophysics (739); 
Time domain astronomy (2109)
}

\section{Introduction}
\label{sec:intro}

All neutron stars are born as hot, differentially rotating proto-neutron stars (PNS), which radiate their gravitational binding energy via thermal neutrino emission over a timescale of seconds (e.g., \citealt{Burrows&Lattimer86,Pons+99,Roberts12}; see \citealt{Roberts&Reddy17} for a recent review).  Most PNS are formed following the core collapse and supernova explosion of a massive star, as evidenced in one important case by the $\sim 13\,{\rm s}$ neutrino burst observed from supernova 1987a \citep{Hirata+87,Bionta+87}.  However, a similarly hot and massive PNS-like remnant can form from the merger of a binary neutron star (BNS) system (e.g., \citealt{Dessart+09,Ciolfi+17,Fujibayashi+20,Sumiyoshi+21}), in those cases when the remnant does not promptly collapse into a black hole (e.g., \citealt{Shibata&Taniguchi06}).   

Rapid rotation, if present, can play a key role in the evolution and fate of PNS.  For example, rotation affects the oscillation frequencies of the remnant (e.g., \citealt{Ferrari+04}) and may be responsible for generating a strong magnetic field through linear winding of the initial field (e.g., \citealt{Shibata+21,Palenzuela+21}), turbulence generated by the Kelvin-Helmholtz (e.g., \citealt{Zrake&MacFadyen13,Kiuchi+15}), Rayleigh-Taylor (e.g., \citealt{Skoutnev+21}) or magneto-rotational (e.g., \citealt{Akiyama+03,Moesta+15}) instabilities, or via an $\alpha-\Omega$ dynamo (\citealt{Thompson&Duncan93,Raynaud+20,White+22}).  Rapid and differential rotation also provides a source of free energy, which---if efficiently tapped---could help power the supernova explosion (e.g., \citealt{Thompson+05}) or outflows from the BNS merger remnant (e.g., \citealt{Fujibayashi+20})

In BNS mergers, the centrifugal support provided by rotation is crucial to the stability of the remnant against gravitational collapse and hence its expected lifetime following the merger (see \citealt{Bernuzzi20,Sarin&Lasky21} for recent reviews).  The maximum stable mass of a cold non-rotating neutron star is given by the Tolman-Oppenheimer-Volkoff (TOV) limit, $M_{\rm TOV}$, its value being a defined property of the neutron star equation of state (EOS).  Solid body rotation can act to increase the effective value of $M_{\rm TOV}$ by up to $\approx$ 20\% for the maximum allowed angular momentum corresponding to the mass-shedding limit (e.g., \citealt{Baumgarte+00}).  This opens the possibility that mergers involving low-mass neutron star binaries may leave {\it supramassive} neutron star (SMNS) remnants (e.g., \citealt{Paschalidis+12,Kaplan+14}), which could survive for minutes to hours or longer after the merger, before losing sufficient angular momentum (e.g., as a result of magnetic dipole braking) to collapse into a black hole.  Due to their powerful neutrino luminosity, expected strong ($\sim$ magnetar-like) magnetic fields and prodigious reservoirs of rotational energy $E_{\rm rot} \sim 10^{52}-10^{53}$ erg, such merger remnants could strongly impact the post-merger electromagnetic counterparts (e.g., \citealt{Metzger+08,Bucciantini+12,Giacomazzo&Perna13,Gao+13,Metzger&Piro14,Gompertz+14,Sarin+22}).  

Although BNS merger remnants are typically formed with enough total angular momentum to generate a SMNS rotating near breakup (e.g., \citealt{Radice+18}), the remnant is not rotating as a solid body.  Numerical relativity simulations show that the merger process generates a slowly rotating core surrounded by a hot quasi-Keplerian envelope (e.g., \citealt{Kastaun+16,Kastaun+17,Ciolfi+17,Ciolfi+19,Kastaun&Ohme21}).  Depending on how mass and angular momentum are transported radially within the star over longer, secular timescales (ranging from tens of milliseconds to as long as seconds), a remnant may remain stable as it evolves into a solid body SMNS state (e.g., \citealt{Fujibayashi+20}), or it may become unstable and undergo dynamical collapse.  For example, if the remnant core were to remain slowly rotating while most of the angular momentum were removed from the outer layers (e.g., by a viscously expanding disk and associated outflows), the remnant could in principle be driven to collapse, even if its mass permits a SMNS solution (e.g., \citealt{Beniamini&Lu21}).  In part as a result of this uncertainty, the BNS properties (total mass, mass ratio) which result in black hole versus SMNS formation for a given EOS remain poorly understood.  Nevertheless, the stakes are high: the ability to delineate this boundary can be used to place an upper bound on $M_{\rm TOV}$ in events where the binary mass is measured through gravitational wave observations but SMNS formation is excluded by electromagnetic observations (e.g., \citealt{Margalit&Metzger17,Shibata+17,Rezzolla+18}; see also \citealt{Lawrence+15,Fryer+15}).

The secular timescale internal angular momentum evolution of a differentially rotating PNS remnant depends on the mechanisms of angular momentum transport throughout the star at different times in its evolution.  In outer regions of the star, where angular velocity $\Omega(r)$ typically decreases with radius, the magneto-rotational instability (MRI) can grow on the dynamical timescale (e.g., \citealt{Balbus&Hawley91,Akiyama+03,Wheeler+15}).  Magnetohydrodynamical (MHD) simulations of BNS merger remnants find that the MRI transports angular momentum efficiently through Reynolds and Maxwell stresses, corresponding to an effective dimensionless viscosity parameter $\alpha \gtrsim 10^{-2}$ (e.g., \citealt{Kiuchi+18}).  Viscous neutrino-radiation hydrodynamics simulations demonstrate how the differentially rotating remnant transforms into a SMNS for an $\alpha$-viscosity with a globally constant value of $\alpha$ \citep{Fujibayashi+20}.  

On the other hand, near the center of the star $d\Omega/dr > 0$ and hence the MRI will not operate.  Thus, an alternative source of angular momentum transport is needed to bring the core into solid body rotation.  One such possible source of transport is neutrino viscosity, which operates while the PNS is opaque to neutrinos (e.g., \citealt{Guilet+15}).  Other sources of ``magnetic'' viscosity can operate in addition to the MRI.  In the immediate aftermath of the merger, turbulence generated by the Kelvin-Helmholtz (e.g., \citealt{Zrake&MacFadyen13,Kiuchi+18}) or Rayleigh-Taylor instabilities (e.g., \citealt{Skoutnev+21}) acts initially to amplify the small-scale magnetic field to large values $\gtrsim 10^{16}$ G.  Magnetic winding will also amplify the toroidal magnetic field in the PNS core (e.g., \citealt{Palenzuela+21,Shibata+21,Aguilera-Miret+22}); however, without a physical mechanism to regenerate a poloidal magnetic field from the amplified toroidal field, no radial Maxwell stress can result.  A promising candidate for such a coupling mechanism is the Spruit-Tayler dynamo \citep{Spruit02,Fuller+19}.

Here we present a preliminary exploration of the various angular momentum processes which can operate in a differentially rotating millisecond PNS, considering the application to both core collapse supernovae and BNS mergers.  In Section \ref{sec:Jtransport}, we estimate, as a function of radius and time after PNS birth, the effective strength of the viscosity arising from different physical processes.  We do this by means of analytic estimates, employing a one-dimensional time-dependent PNS cooling model to motivate the time-evolving background state.  Although the MRI can transport angular momentum efficiently in the outer layers of the star, even in the core we find that the combination of 
convection and the Spruit-Tayler dynamo operate efficiently on the cooling age.  Thus, we predict that the cores of rapidly rotating PNS will quickly enter a state of solid body rotation, faster than angular momentum can be added or removed from the quasi-Keplerian outer envelope/disk.  

Thus motivated, in Section \ref{sec:model} we explore further the implied fate of BNS merger remnants by developing a two-zone (solid-body rotating core + quasi-Keplerian envelope/disk) toy model for the long-term evolution and stability of the merger remnant as the core accretes mass and angular momentum from the envelope.  We use this model to delineate the initial core properties that result in an accretion-induced collapse versus survival as a long-lived SMNS remnant.  Our results illustrate that the SMNS boundary may not be defined exclusively in terms of the remnant mass, thus complicating the use of electromagnetic observations which rule out SMNS formation to place an upper limit on the TOV mass.  We summarize and discuss our conclusions in Section \ref{sec:conclusions}.


\section{Angular Momentum Transport in Rotating PNS}
\label{sec:Jtransport}

\subsection{Models for Cooling PNS in Supernovae and BNS Mergers}
\label{sec:PNSmodel}

We employ a $1.6 \, M_{\odot}$ PNS stellar evolution model similar to the ones described in \citet{Roberts&Reddy17}. This cooling model is produced by following the collapse and bounce of the $15 \, M_\odot$ progenitor of \citet{Woosley+2002} in spherical symmetry until the shock reaches a baryonic mass coordinate of 1.6$M_{\odot}$. At this point, the material from the shock outward is excised from the grid and replaced by an outer boundary condition. The cooling of the remnant PNS is then followed for $\sim 100 \,\textrm{s}$. Neutrino transport in the star is followed using a multi-group, two-moment scheme that is closed via an approximate solution to the static Boltzmann equation using characteristics \citep{Roberts12}. Three flavors of neutrinos are followed, $\nu_e$, $\bar \nu_e$ and $\nu_x = \{\nu_\mu, \bar\nu_\mu, \nu_\tau, \bar \nu_\tau \}$ each using twenty logarithmically spaced energy groups. We employ the Lattimer-Swesty equation of state with an incompressiblity parameter $K = 220 \, \textrm{MeV}$ \citep{Lattimer&Swesty91} using the tables described in \cite{Schneider+2017}. The neutrino opacties are taken from \cite{Burrows+2006}. The effects of convection are not included in the models.

The models provide the radial profiles of the following quantities as a function of time since core bounce: mass density $\rho$, enclosed mass $M_{\rm enc}$, energy flux carried by neutrinos $F$, temperature $T$, pressure $P$, electron fraction $Y_e$, entropy $s$, lepton fraction $Y_l$, and adiabatic index $\gamma_s$.  The models also provide the energy densities, $e_{\nu}$, and local mean free-paths, $\lambda_{\nu}$, as a function of neutrino energy $\epsilon$ for each of the three neutrino species.

Our estimates of the Spruit-Tayler dynamo require the Brunt-V\"ais\"al\"a frequency $N$,
\be N^{2} = \frac{g}{\gamma_s} \left(\frac{d\ln P}{d\ln n}-1\right)\left(\frac{d\ln n}{dr}\right),
\label{eq:Brunt}
\ee
where $g = GM_{\rm enc}/r^{2}$ is the gravitational acceleration in the Newtonian approximation, $n$ is the baryon number density, and $\gamma_s$ is the adiabatic index.

Although our goal is to study mechanisms of angular momentum transport in PNS, we do not include rotation effects in the stellar evolution models.  This is primarily due to the lack of a self-consistent way to include rotation and its time-evolution in a one-dimensional model without significant modifications to the available infrastructure \citep{Roberts&Reddy17} as well as the explorative ($\sim$order of magnitude) nature of our estimates.  Including rotation would have a number of quantitative effects on the stellar evolution, such as: reducing the effective gravity, particularly in the outer layers of the star; affecting the criterion for convection (e.g., \citealt{Tassoul78}); providing an additional source of heating from viscous dissipation (e.g., \citealt{Thompson+05}); and the potential for additional mixing processes driven by rotation (e.g., \citealt{Heger+00,Aerts+19} and references therein). 

We begin our calculations following core bounce, after the outer mass of the star has been removed from the grid to approximate the effects of a successful supernova explosion (see \citealt{Roberts&Reddy17}).  Although our calculation is specific to the core collapse case, apart from the somewhat different mass scale and initial lepton number distribution, we expect the thermal evolution of a BNS merger remnant to qualitatively match that of a PNS formed from a core collapse supernova.  For example, regarding the initial entropy distribution, while the outer mantle of a PNS formed during core collapse is heated by the outwards propagating shock generated at core bounce (e.g., \citealt{Janka+07}), the outer layers of a BNS merger remnant are also preferentially heated during the merger process (e.g., \citealt{Raithel+21}).  Supporting a similar evolution, the neutrino luminosities of neutron star merger remnants, as calculated using axisymmetric 2D multigroup flux-limited-diffusion radiation-hydrodynamics simulations \citep{Dessart+09} are similar, at least up to timescales of a few hundred milliseconds after the merger, with those of PNS formed in core-collapse supernovae at the same epoch after core bounce (e.g., \citealt{Pons+99,Roberts+12}).  In the merger case, we define the $t = 0$ point as when the PNS has contracted to a radius $\approx 20$ km, motivated by the approximate size of the remnant predicted by numerical relativity merger simulations (e.g., \citealt{Ciolfi+17}).

Although our PNS models do not include rotational effects, the angular momentum transport processes do depend on the angular velocity profile $\Omega(r)$ of the star.
We consider two scenarios for $\Omega(r)$, representative of core collapse supernovae and BNS mergers, respectively.  
For the supernova case, 
we follow the prescription of \citet[their Eq.~4]{Thompson+05} and assume that the iron core is rotating 
as $\Omega(r) = \Omega_0/\left[1+(r/R_\Omega)^2\right]$ prior to collapse.
We then conserve the specific angular momentum $j = r^{2}\Omega$ within mass shells during the collapse.  
This rotation profile implies an iron core that is roughly solid-body rotating out to $r = R_\Omega$. We choose $R_\Omega = 1000\,{\rm km}$ and $\Omega_0 = 2\pi/(2\,{\rm s})$ as these reasonably approximate a rotating stellar progenitor model from \citet[see Fig.~1 in \citealt{Thompson+05}]{Heger+00}. This choice also ensures that $\Omega(r) \lesssim \Omega_{\rm K}$ at all times of interest, where $\Omega_{\rm K} = (GM_{\rm enc}/r^{3})^{1/2}$ is the Keplerian angular velocity.  

\begin{figure}
    \centering
    \includegraphics[width=0.5\textwidth]{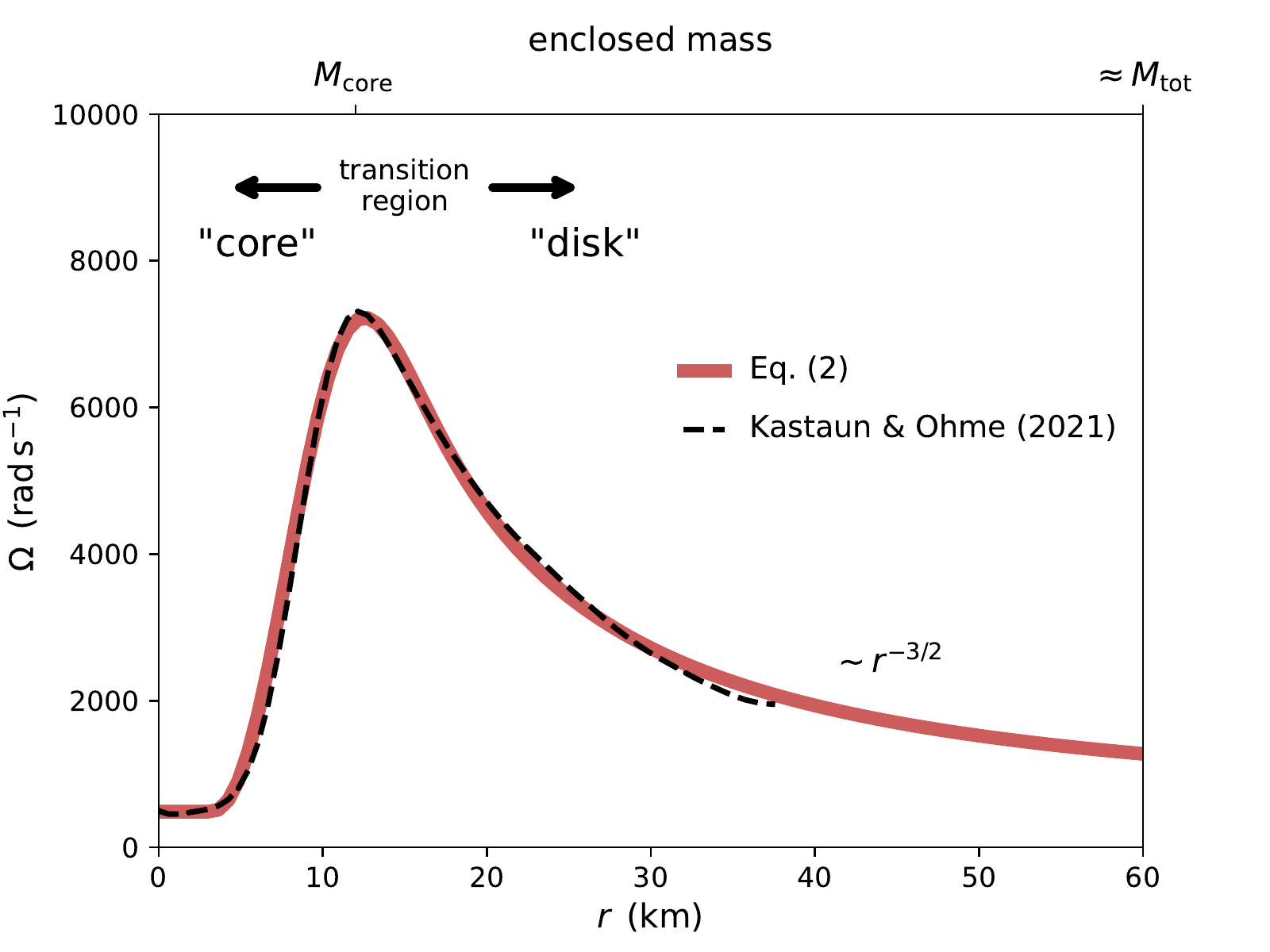}
    \caption{Schematic of BNS merger remnants as considered in this work. The rotational profile of post-merger remnants has been found by numerical relativity simulations to be qualitatively similar to that shown above. Black dashed curves show the profile reported by \cite{Kastaun&Ohme21} (after removing the frame-dragging effect), and the solid-red curve shows the analytic approximation employed in generating Fig.~\ref{fig:time_radius_remnant} (Eq.~\ref{eq:OmegaBNS}).  The system can be thought of as composed of an initially slowly-rotating core at small radii that is surrounded by a $\sim$centrifugally-supported envelope/disk at larger distances. The mass contained within the core, $M_{\rm core}$ is typically a large fraction of the total merger-remnant mass $M_{\rm tot}$ (with current numerical relativity merger simulations suggesting $M_{\rm core}/M_{\rm tot} \sim 0.8$).}
    \label{fig:schematic}
\end{figure}

For the BNS merger case, we instead adopt a rotational profile motivated by the findings of numerical relativity simulations (e.g., \citealt{Uryu+17,Kiuchi+18,Kastaun&Ohme21}).  In particular, we take:
\begin{equation}
    \Omega(r) =  
    \begin{cases}
    \Omega_{\rm i} + \Omega_0\left(\frac{r}{r_0}\right)^{-3/2}\left[ 1-e^{-\left(\frac{r-r_{\rm i}}{r_0}\right)^{3}} \right] &, r > r_{\rm i} \\
    \\
    \Omega_{\rm i} &, r \leq r_{\rm i} 
    \end{cases}
\label{eq:OmegaBNS}
\end{equation}
We define $\Omega_0$ such that $\Omega \approx \Omega_{\rm K}$ for $r \gg r_0$ and take $r_i = 3$ km, $r_0 = 8$ km and $\Omega_{\rm i} = 0.03\Omega_0$, as we find these reasonably reproduce the $\Omega(r)$ profile found by \citet{Kastaun&Ohme21}, once the frame-dragging contribution has been removed (Figure \ref{fig:schematic}).

\subsection{Sources of Viscosity}
\label{sec:viscosity}

We consider four mechanisms of angular momentum transport:
\begin{enumerate}
    \item Neutrino viscosity
    \item MRI
    \item Convection
    \item Spruit-Tayler dynamo
\end{enumerate}
We model each of these as a viscous process with an associated kinematic viscosity $\nu$ and viscous timescale
\be
t_{\rm visc} = \frac{r^{2}}{\nu}.
\label{eq:tvisc}
\ee

In optically thick regions of the star (valid throughout most of the remnant mass on timescales of interest), the neutrino viscosity can be written as (\citealt{vandenHorn&vanWeert84,Thompson+05,Guilet+15})
\begin{equation}
\label{eq:nunu}
\nu_{\nu} 
= \sum_{\nu_i \in \left\{\nu_e, ... \right\}}  \frac{4}{15} \frac{e_{\nu_i}}{\rho c} \left\langle \lambda^*_{\nu_i} \right\rangle
\end{equation}
where $e_{\nu_i}$ is the energy density of neutrino species $\nu_i$ and $\left\langle \lambda^*_{\nu_i} \right\rangle$ is an appropriately averaged neutrino mean free path defined in Eq.~(\ref{eq:Appendix_average_mfp}).
Note that {\it $\left\langle \lambda^*_{\nu_i} \right\rangle$ is not an energy-weighted mean free path}, and instead involves a more complicated weighting scheme. This has led to some confusion in the published literature, and we discuss this point further in Appendix~\ref{sec:Appendix}.

We approximate the effective viscosity due to MRI turbulence as (e.g., \citealt{Akiyama+03})
\be \nu_{\rm MRI} = \alpha h_{\rm t}^2 \Omega q, \,\,\,\,d\Omega/dr < 0 
\label{eq:nuMRI}
\ee 
where $\Omega$ is the angular velocity, $q \equiv d\ln\Omega/d\ln r$ is the dimensionless shear, $\alpha$ is a dimensionless parameter that accounts for the strength of the Maxwell and shear stresses relative to the pressure.  We take $\alpha = 0.1$ in our calculations, though its true value could be an order of magnitude smaller (e.g., \citealt{Kiuchi+18}).
We evaluate the stability criterion for the MRI following~\citet{Menou_2004,Balbus&Hawley91}, where a necessary condition for stability is
\begin{align}
    \frac{d\Omega^2}{d\ln r} > 0,
\end{align}
and we deem the MRI to be active whenever this condition is not satisfied.
Note that we do not include any buoyancy terms in this criterion because in regions of the star where rotation is most significant (near-Keplerian) the PNS is heavily deformed and the Brunt-V\"ais\"al\"a frequency is reduced relative to that in a non-rotating star (our assumed background state).

The length-scale $h_{\rm t}$ entering Eq.~(\ref{eq:nuMRI}) is the characteristic coherence length of the turbulence.  For a slowly rotating star with a sharp outer boundary, \citet{Thompson+05} suggest to take $h_{\rm t}$ equal to the radial pressure scale-height $h \equiv (d{\rm ln}P/dr)^{-1}$ (e.g., \citealt{Thompson+05}).  However, again, in our models for PNS rotating near break-up ($\Omega \simeq \Omega_{\rm K}$), with surface layers akin to a vertically thick accretion disk (or ``decretion'' disk), we instead take the turbulent length scale to be the disk vertical scale-height, $h_{\rm t} \sim r/3$.

Regions of the PNS become unstable to convection, due to entropy and/or leptonic composition gradients (e.g., \citealt{Burrows&Lattimer88,Mezzacappa+98,Roberts+12,Nagakura+20}).  Convection can transport angular momentum via mixing following fluid advection.  We approximate the effective viscosity due to convection as,
\be \nu_{\rm conv} \approx v_{\rm c} h, \ee
where $v_{\rm c} \approx(F_{\rm conv}/\rho)^{1/3}$ is the convection speed expressed in terms of the convective flux $F_{\rm conv}$. For simplicity we assume that the convective flux everywhere equals the neutrino flux at the surface of the model.  Although this prescription is crude it captures the relevant scaling in the efficient convection regime.  

Even in stably stratified regions of the star, magnetic field amplification arises due to winding by differential rotation.  This growth saturates in an instability, which results in the generation of a poloidal magnetic field component that couples with the azimuthal field to generate a radial Maxwell stress via the `Spruit-Tayler' (TS) dynamo (e.g., \citealt{Spruit02}).
The strength of the effective viscosity arising from this coupling depends on the saturation amplitude of the instability.  Here we employ the saturation prescription of~\citet{Fuller+19}, which gives a viscosity
\be \nu_{\rm TS} \approx r^2 \frac{\Omega^3}{N^{2}}.
\label{eq:nuTS}
\ee
This choice is motivated by its success in matching observations of red giant core rotation (e.g., \citealt{Fuller+19}), though we note that other prescriptions yield a substantially weaker dynamo and hence lower viscosity, typically by factors of $(\Omega/|N|)^2 \sim 10^{-1}-10^{-2}$~\citep{Spruit02}. 

We turn off the TS viscosity in convectively unstable regions because we are not sure how this dynamo process interacts with convective instabilities, and in any case convection is always fast enough to ensure near-uniform rotation where it acts. Similar uncertainties apply to the interaction of other angular momentum transport processes.  For instance, a region which is MRI turbulent would presumably also suppress the TS instability (and vice versa).  Likewise, neutrino viscosity can suppress the MRI in the core of the PNS depending on the magnetic field strength (e.g., \citealt{Guilet+15}).  Nevertheless, these uncertainties are mostly irrelevant as long as we focus our consideration on the dominant source of viscosity (shortest viscous timescale) on any radial scale.

Finally, note that there exist other potential sources of angular momentum transport that we neglect, for instance by waves excited in the PNS interior (e.g., \citealt{Fuller+14,Gossan+20}), pumping due to convective overshoot between adjacent radiative and convective layers (\citealt{Kissin&Thompson18}), 
hydrodynamic instabilities within radiative regions (e.g., \citealt{Zahn74,Prat+16}) or boundary layers \citep{Belyaev&Rafikov12,Belyaev&Quataert18},
or spiral density waves induced by non-axisymmetric deformations of the merger remnant \citep{Goodman&Rafikov01,Rafikov16,Nedora+19,Nedora+21}.
These processes could also be important but are more difficult to quantify in comparison to those sources we have focused on. 

\subsection{Analytic Estimates}

Before proceeding to our numerical results, we provide analytic estimates for the effective viscous timescales associated with different processes.  When necessary to specify, we consider a ``typical'' location at the radius $r \sim R/2$ within the remnant of radius $R \sim 20$ km and mass $M_{\rm rem} \approx 2.6M_{\odot}$, where the angular rotational velocity $\Omega \lesssim \Omega_{\rm K} = (GM_{\rm rem}/R^{3})^{1/2}$.

In Appendix~\ref{sec:Appendix} we derive an approximate analytic expression for the neutrino viscosity (Eq.~\ref{eq:nunu}) as a function of density, temperature, and neutrino degeneracy $\eta_{\nu_i} \equiv \mu_{\nu_i} / kT$ (where $\mu_{\nu_i}$ is the chemical potential of neutrino species $\nu_i$), by
considering six species of neutrinos in local thermodynamic equilibrium and assuming that the opacity is dominated by scattering onto non-degenerate neutrons and protons (Eq.~\ref{eq:Appendix_nunu_final}),
\begin{eqnarray}
\label{eq:nunu_full}
  \nu_\nu 
  \approx 1.66 \times 10^{10} &\,& {\rm cm}^2 \, {\rm s}^{-1} \, 
  \left(\frac{T}{10 \, \textrm{MeV}}\right)^2 \left(\frac{\rho}{10^{13} \, {\rm g \, cm}^{-3}}\right)^{-2}
  \nonumber\\
  &\times&
  \frac{f(Y_e)}{0.7}
  \left[ 1 +
  \frac{\eta_{\nu_e}^2 + \eta_{\nu_\mu}^2 + \eta_{\nu_\tau}^2}{\pi^2} 
  \right]
  .
\end{eqnarray}
Here $f(Y_e)$ is a weak function of the electron fraction $Y_e$, that spans the limited range $f(Y_e) \in [0.68,0.74]$ for $Y_e \in [0,0.5]$ (see Eq.~\ref{eq:Appendix_fYe}).
The expression above is analogous to the \cite{Keil+96} result that is often used in the literature (e.g.,~\citealt{Guilet+15,White+22}), but generalized to arbitrary neutrino degeneracy and with an updated numerical value.
Although mu and tau neutrinos remain non-degenerate in our context,
electron neutrino degeneracy can become appreciable within the PNS or merger-remnant core ($\eta_{\nu_e} \sim 10$; Eq.~\ref{eq:Appendix_eta}). A convenient analytic expression applicable in the non-degenerate and highly-degenerate regimes of interest is therefore
\begin{eqnarray}
\label{eq:nunu_cases}
    \nu_\nu 
    &\approx&
    10^{10} \, {\rm cm}^2 \, {\rm s}^{-1} 
    \\ \nonumber
    &\,&\times
    \begin{cases}
    1.7 \, \left(\frac{T}{10 \, \textrm{MeV}}\right)^2 \left(\frac{\rho}{10^{13} \, {\rm g \, cm}^{-3}}\right)^{-2}
    &,\, \eta_{\nu_e} \approx 0
    \\
    7.1 \,
    \left(\frac{\rho}{10^{13}\,{\rm g\,cm}^{-3}}\right)^{-4/3}
    \left(\frac{Y_{\nu_e}}{0.1}\right)^{2/3}
    &,\, \eta_{\nu_e} \gg 1
    \end{cases}
\end{eqnarray}
where $Y_{\nu_e}$ is the electron neutrino number fraction. Since neutrino degeneracy can only increase the viscosity, one can effectively take the relevant viscosity as the maximum of the two cases above. 
Scaling the density to the mean value $\bar{\rho} \sim 3M_{\rm rem}/4\pi R^3 \approx 1.5\times 10^{14}$ g cm$^{-3}$ and adopting the non-degenerate case, the neutrino viscosity timescale can thus be written
\be
t_{\rm \nu} \sim \frac{r^{2}}{\nu_{\nu}} \sim 600\,{\rm s}\,\left(\frac{T}{50\,{\rm MeV}}\right)^{-2}\left(\frac{\rho}{\bar{\rho}}\right)^{2},
\label{eq:tnu}
\ee
where we have scaled the temperature $T$ to a typical peak value $\sim 50$ MeV in the post-merger remnant (e.g., \citealt{Perego+19}).
A similar estimate is obtained if one instead assumes that electron neutrinos are degenerate.
This timescale is much longer than the cooling evolution time, suggesting neutrino viscosity is irrelevant throughout the bulk of the star.  

The timescale for angular momentum transport due to MRI turbulence is approximately given by
\be
t_{\rm MRI} \sim \frac{r^{2}}{\alpha h_{\rm t}^{2}\Omega} \sim 5\times 10^{-3}\,{\rm s}\, \left(\frac{\alpha}{0.1}\right)^{-1}\left(\frac{\Omega}{\Omega_{\rm K}}\right)^{-1}\left(\frac{h_t}{r/3}\right)^{-2},
\label{eq:tMRI}
\ee
where we have used Eq.~(\ref{eq:nuMRI}) with $q = 1$.

In regions where convection carries the outwards luminosity of the PNS, the convective luminosity $L_{\rm c} \sim 4\pi r^{2}\rho v_{\rm c}^{3}$ must approximately equal the surface neutrino luminosity $L_{\nu}$, where $\rho$ is the stellar density and $v_{\rm c}$ the convective velocity.  This gives for the latter,
\begin{eqnarray}
v_{\rm c} &\sim& \left(\frac{L_{\nu}}{4\pi r^{2}\rho}\right)^{1/3} 
\\ \nonumber
&\approx& 4\times 10^{8}\,{\rm cm\,s^{-1}}\,\left(\frac{L_{\nu}}{10^{53}\,{\rm erg\,s^{-1}}}\right)^{1/3}\left(\frac{\rho}{\bar{\rho}}\right)^{-1/3},
\label{eq:vc}
\end{eqnarray}
where we have normalized $L_{\nu}$ to a characteristic value $\sim 10^{53}$ erg s$^{-1}$ for a PNS or merger remnant on timescales of hundreds of milliseconds after birth (e.g., \citealt{Pons+99,Dessart+09}).  This gives a characteristic viscous timescale due to convection,
\begin{eqnarray}
t_{\rm conv} &\sim& \frac{r^{2}}{\nu_{\rm conv}} \sim \frac{r^{2}}{h v_{\rm c}} 
\\ \nonumber
&\sim& 3\times 10^{-2}\,{\rm s}\,\left(\frac{h}{r/10}\right)^{-1} \left(\frac{L_{\nu}}{10^{53}\,{\rm erg\,s^{-1}}}\right)^{-1/3},
\label{eq:tconv}
\end{eqnarray}
where we have used Eq.~(\ref{eq:vc}) with $\rho = \bar{\rho}$.  

Finally, using Eq.~(\ref{eq:nuTS}) the TS timescale is given by \begin{eqnarray}
t_{\rm TS} &\sim& \frac{r^{2}}{\nu_{\rm TS}} \sim \frac{N^{2}}{\Omega^{3}} \sim \frac{\Omega_{\rm K}^{2}}{\Omega^{3}}\left(\frac{h}{r}\right)^{-1} \nonumber \\
&\sim& 5\times 10^{-4}\,{\rm s}\,\left(\frac{h}{r/10}\right)^{-1}\left(\frac{\Omega}{\Omega_{\rm K}}\right)^{-3},
\label{eq:tTS}
\end{eqnarray}
where in the third equality we have approximated the Brunt-V\"ais\"al\"a frequency (Eq.~\ref{eq:Brunt}) as $N \simeq (g/h)^{1/2}$ with $g = GM_{\rm rem}/r^{2}$.  If one were to adopt the \citet{Spruit02} saturation criterion instead of \citet{Fuller+19}, then $t_{\rm TS}$ would be larger by a factor $(|N|/\Omega)^{2} = (\Omega/\Omega_{\rm K})^{-2}(r/h) \sim 10(\Omega/\Omega_{\rm K})^{-2} \sim 10-100$.

In summary, in outer regions of the star rotating at or near break-up ($\Omega \sim \Omega_{\rm K}$) where the MRI operates, we have $t_{\rm TS} \lesssim t_{\rm MRI} \lesssim t_{\rm conv} \ll \tau_{\rm KH}$, where
\be
\tau_{\rm KH} \sim \frac{GM_{\rm rem}^{2}}{RL_{\nu}} \sim 9\,{\rm s} \,\,\left(\frac{L_{\nu}}{10^{53}\,{\rm erg\,s^{-1}}}\right)^{-1}
\label{eq:tauKH}
\ee
is the Kelvin-Helmholtz timescale (roughly equal to the thermal timescale over which the PNS properties change at a characteristic interior point). 
In the more slowly rotating core ($\Omega < \Omega_{\rm K})$, even if the MRI is inactive, we can have $t_{\rm conv}, t_{\rm TS} < \tau_{\rm KH}.$  As we now discuss, these estimates are confirmed by our more detailed numerical models.      

\subsection{Numerical Results}

\begin{figure*}
    \centering
    \includegraphics[width=0.65\textwidth]{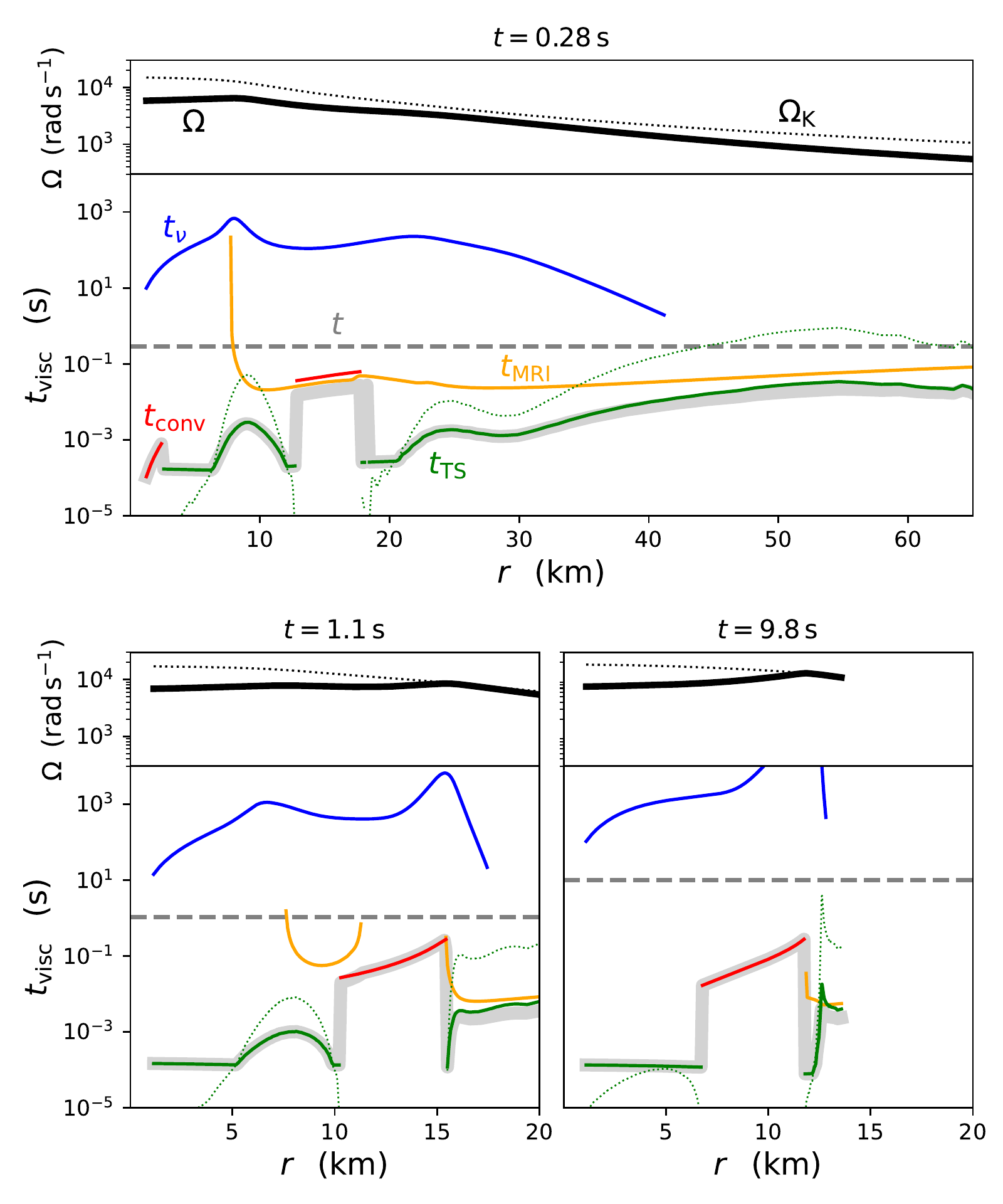}
    \caption{Timescale for angular momentum transport $t_{\rm visc}$ (Eq.~\ref{eq:tvisc}) within a PNS remnant formed from the core collapse of a rotating massive star as a function of radius at several snapshots in time measured from core bounce, as marked above each plot. The imposed angular velocity profile $\Omega(r)$ is shown as a solid black curve in the top panel of each snapshot, and can be compared to the Keplerian angular velocity $\Omega_{\rm K}$ (dotted black).  We assume $\Omega(r) = (2\pi/2\,{\rm s})/\left[1+(r/1000\,{\rm km})\right]$ in the iron core prior to collapse and that specific angular momentum is conserved on mass shells during collapse. 
    We show viscous times separately for each of the angular momentum transport processes we consider in \S\ref{sec:viscosity}: neutrino viscosity (blue; limited to optically-thick regions), MRI (yellow), convection (red), and the TS dynamo (green; solid---nominal case adopting the \citealt{Fuller+19} saturation condition; dotted---using the \citealt{Spruit02} saturation prescription instead).
    The total viscous time arising from the sum of the viscosities is shown with the solid grey curve.
    The total viscous time is shorter than the PNS's thermal timescale $\gtrsim$ its age $\sim t$ (horizontal dashed grey curve), suggesting that the object quickly comes into solid body rotation.}
    \label{fig:time_radius_cc}
\end{figure*}

\begin{figure*}
    \centering
    \includegraphics[width=1.0\textwidth]{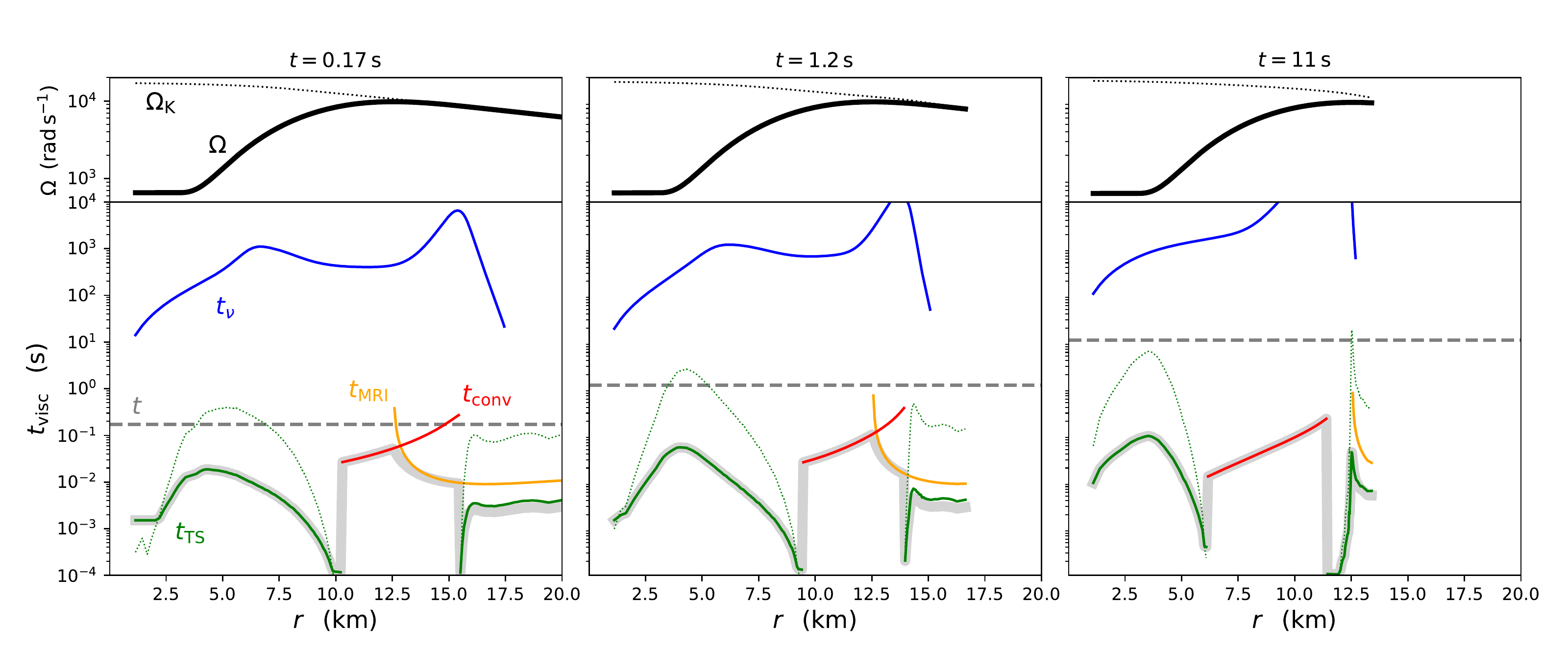}
    \caption{Same as Fig.~\ref{fig:time_radius_remnant}, but calculated for an imposed angular velocity profile appropriate to a BNS merger remnant (Eq.~\ref{eq:OmegaBNS}).}
    \label{fig:time_radius_remnant}
\end{figure*}

Figures \ref{fig:time_radius_cc} and \ref{fig:time_radius_remnant} present the radial profiles from our numerical PNS cooling calculations (\S\ref{sec:model}) of the angular momentum redistribution timescale $t_{\rm visc}$ (Eq.~\ref{eq:tvisc}) associated with each source of viscosity (\S\ref{sec:viscosity}) in the case of supernovae and BNS mergers, respectively. The different panels in each figure show different snapshots in the cooling evolution of the PNS.  Solid grey curves show the total viscous time, calculated using the sum of all sources of viscosity.  Black solid and black dotted curves at the top of each panel show the assumed profiles of $\Omega$ and $\Omega_{\rm K}$, respectively, as described in Section \ref{sec:PNSmodel}.

Two main questions we want to address are: 
\begin{enumerate} 
\item{Will the supernova or merger remnant enter approximate solid body rotation over timescales shorter than the remnant's thermal properties (and hence all of the non-MRI viscosity sources) are evolving?}
\item{Can angular momentum be transported into the core as fast as mass is being supplied from the outer rapidly-rotating layers or quasi-Keplerian disk?}
\end{enumerate}

Question (1) is addressed by comparing $t_{\rm visc}$ to the age $\sim$cooling time of the PNS, $t$ (shown as dashed grey horizontal curves for each snapshot in Figs.~\ref{fig:time_radius_cc}, \ref{fig:time_radius_remnant}).  
In both the supernova and BNS merger cases, neutrino viscosity is generally the least effective, insofar that $t_{\nu} \gg t$.
However, the combination of convection and/or the TS dynamo are efficient throughout all epochs we consider ($t_{\rm conv}, t_{\rm TS} \ll t$), even in regions of the star where $d\Omega/dr > 0$ and the MRI is inactive.  For example, in the first tens of milliseconds of the merger case, we have $t_{\rm TS}, t_{\rm conv} \sim 1-100\,{\rm ms}$ across radii $r \lesssim 10\,{\rm km}$.  We therefore expect that regions with $\Omega \lesssim \Omega_{\rm K}$ will be brought into solid body rotation on a similar timescale.  

The insensitivity of our conclusion to whether TS or convection transport operates at any point or epoch is potentially important, because the size and locations of the convective zones are sensitive to details which are either not captured by our simulation (e.g., the effects of rotation or differences in the lepton gradient from the one we have assumed) or are otherwise hard to quantify (e.g., uncertainties in the nuclear symmetry energy; \citealt{Roberts+12}).  On the other hand, there remain uncertainties about whether convection will truly enforce solid-body rotation (see \S\ref{sec:conclusions} for discussion).

Even in regions of the star where the MRI operates, the TS dynamo is seen to compete ($t_{\rm TS} \sim t_{\rm MRI}$), although which of these ``wins'' will be sensitive to their relative saturation strengths.  For the MRI we have assumed an effective viscosity $\alpha = 0.1$, an order of magnitude higher than found by \citet{Kiuchi+18}, and hence our estimate of $t_{\rm MRI}$ could be seen as a lower limit.  On the other hand, adopting the \citet{Spruit02} prescription for the TS saturation could increase $t_{\rm TS}$ by a factor of $(|N|/\Omega)^2 \sim 10-100$.
For comparison purposes, the TS timescale adopting this \citet{Spruit02} prescription is shown as dotted green curves in Figs.~\ref{fig:time_radius_cc},\ref{fig:time_radius_remnant}. Even for this conservative case, the TS timescale is $\lesssim t$ throughout most of neutron star core, lending further strength to the finding that the object comes into solid body rotation over short timescales.

The answer to Question (2) above depends on what mechanisms are responsible for adding mass to the core.  
\citet{Fujibayashi+20} find that neutrino cooling of the outer layers of the remnant allows growth of mass and angular momentum of the central PNS core.
Furthermore, Figs.~\ref{fig:time_radius_cc}, \ref{fig:time_radius_remnant} show that the TS/convective timescale in the stellar interior (where $d\Omega/dr < 0$) is typically comparable or shorter than the MRI timescale in the outer regions ($d\Omega/dr > 0$); this suggests that the disk will be fed from the outside at a rate slower than required for the core to enter solid body rotation.
This conclusion would be further strengthened by the presence of a Keplerian-disk extending to large radii (e.g., from supernova fall-back material or tidal ejecta in the BNS merger case) for which $t_{\rm MRI}$ is even longer than shown in Figs.~\ref{fig:time_radius_cc}, \ref{fig:time_radius_remnant} (comparable to the duration of short gamma-ray bursts $\sim 0.1-1\,{\rm s}$, under the usual assumption that the latter are powered by disk accretion). 

\section{Model for Post-Merger Remnant Evolution}
\label{sec:model}

Motivated by the results of \S\ref{sec:Jtransport}, we now present a toy model for the long-term viscous evolution of post-merger remnants. Though simplified and imprecise in several details, this model paints a new qualitative picture with implications for connecting the lifetime and long-term stability of such remnants to the initial radial distribution of mass and angular momentum established during the dynamical phase of the merger. 

Our model is based on treating the post-merger remnant as a two-zone system comprised of a solid-body rotating neutron star `core' and a surrounding rotationally-supported Keplerian envelope or `disk' (see Fig.~\ref{fig:schematic} for a schematic illustration). This approach is motivated by results of numerical relativity simulations, that generically find the post-merger remnant structure (assuming that prompt-collapse to a black hole has not taken place) to be well characterized by a slowly-rotating cold `TOV-equivalent' core neutron star solution that transitions to a rotationally supported envelope at large radii \citep[e.g.,][]{Kastaun+16,Ciolfi+17,Ciolfi+19,Kastaun&Ohme21}.
Within this two-zone framework, the system's secular evolution is governed by mass and angular momentum transfer due to accretion of material from the disk onto the neutron star core.
As we have shown in \S\ref{sec:Jtransport}, angular-momentum transport within the core is very rapid. This motivates us to consider the limiting case where accreted angular-momentum is instantaneously redistributed within the core, effectively maintaining the core at solid-body rotation.  This limit is reasonable so long as the viscous time throughout the core is shorter than lifetime of the disk, as we have shown is likely the case.

\begin{figure*}
    \centering
    \includegraphics[width=0.7\textwidth]{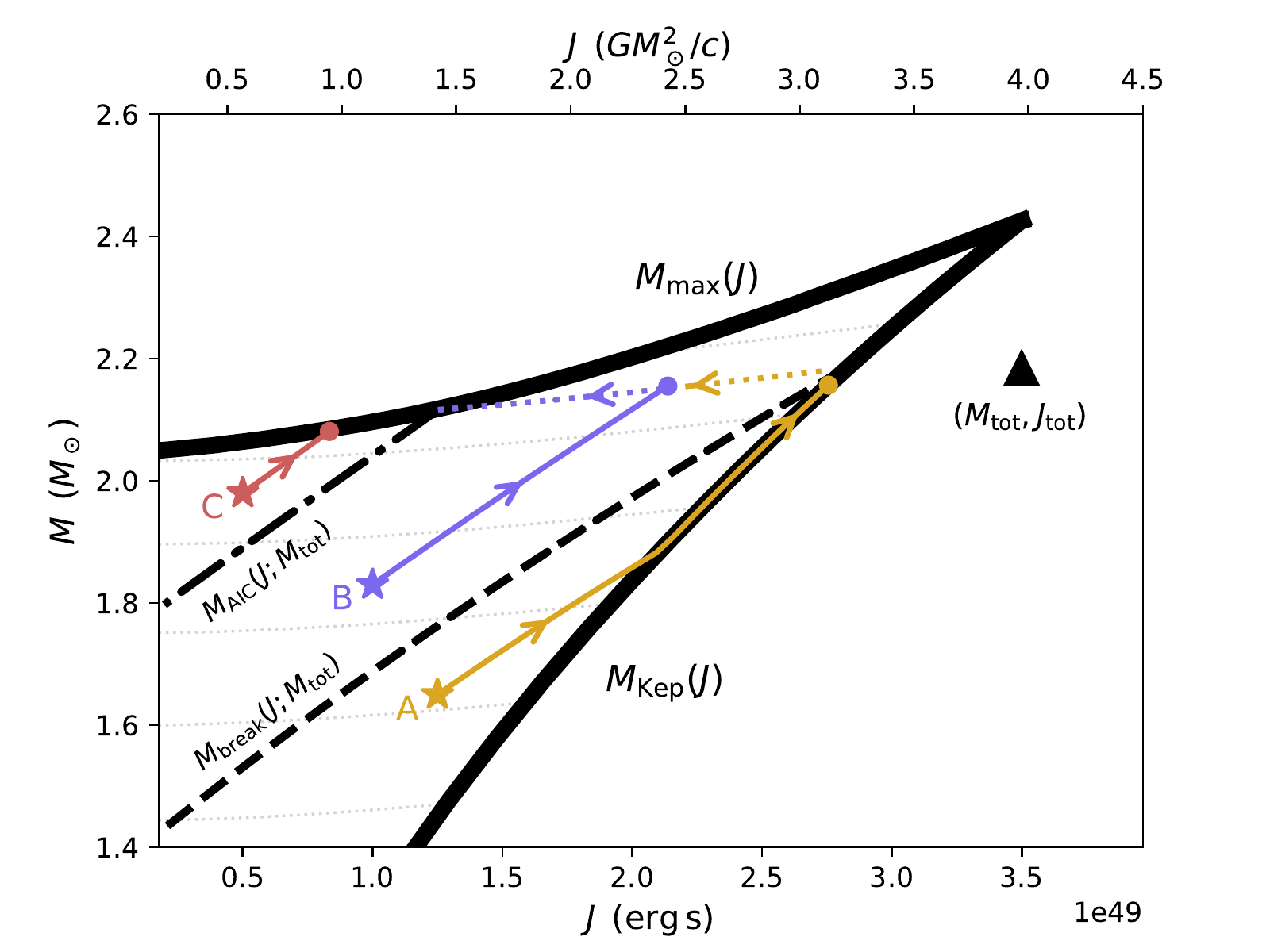}
    \caption{Evolution in the space of mass and angular momentum of the (assumed solid-body rotating) core of a BNS merger remnant under the action of accretion from a surrounding Keplerian envelope/disk.  Solid black curves bound the parameter-space in which solid-body rotating equilibrium neutron star solutions are viable, as set by the maximum mass supported against gravitational collapse $M_{\rm max}(J)$ (top curve) and the mass-shedding limit $M_{\rm Kep}(J)$ (right curve), for an SLy EOS (where $M_{\rm TOV} = 2.04M_{\odot}$).  A black triangle denotes the total remnant mass $M_{\rm tot}$ and angular momentum $J_{\rm tot}$ in an illustrative case residing in the traditionally-considered SMNS regime ($M_{\rm TOV} < M_{\rm tot}  \lesssim 1.2 M_{\rm TOV}$).  Stars show three examples of the initial post-merger core properties which exhibit qualitatively distinct evolution.  Over disk viscous timescales, mass and angular momentum are accreted onto the core, causing it to move up the trajectory illustrated by the appropriately-colored solid curves.  Case B accretes until reaching $M = M_{\rm tot}$, while case A first hits the mass-shedding curve and hence is rotating near break-up when reaching $M = M_{\rm tot}$.  In both Cases A and B, the remnant will only collapse to a black hole once disk accretion has ceased and angular momentum is lost at roughly fixed mass over longer timescales, e.g. via magnetic spin-down (left along the dotted lines).  By contrast, in case C, collapse occurs during the core accretion process itself, bypassing the spin-down process and resulting in very little rotational energy deposited into the remnant environment.  Such an accretion-induced collapse outcome occurs for initial core properties $\{ M_{\rm core},J_{\rm core}\}$ above the critical curve $M_{\rm AIC}(J)$ (dot-dashed black curve; approximately given by Eq.~\ref{eq:M_AIC}). A similar curve, $M_{\rm break}(J)$ (dashed-black curve) separates between cores that are spinning at break-up at the end of viscous evolution and those whose rotation rate is lower (as in cases A, B, respectively). See text for further details.}
    \label{fig:MJ}
\end{figure*}

Under the simplifying conditions above we can gain new insight into the long-term viscous evolution of BNS merger remnants. If angular momentum transport in the core is fast, the core will be well described by a solid-body rotating equilibrium neutron star solution at any point throughout its secular evolution. This evolution slowly increases the core mass and angular momentum due to accretion, driving the core into a new equilibrium solution, or towards instability and collapse.
If we assume a Keplerian rotationally-supported disk that extends down close to the core surface, then the amount of angular momentum gained by the core of mass $M$ and angular momentum $J$ per accreted unit mass is simply 
\begin{equation}
\label{eq:j_Kep}
    \frac{dJ}{dM} 
    \sim j_{\rm Kep}(R_{\rm acc}) = \sqrt{G M R_{\rm acc}},
\end{equation}
the specific angular momentum of accreted material, normalized to that of a Keplerian disk of radius $R_{\rm acc}$; we shall typically set $R_{\rm acc} = \zeta R$, where $R(M,J)$ is the equatorial radius of the neutron star equilibrium solution and $\zeta \sim 1-2$ is an uncertain parameter that depends on the details of the boundary layer separating the core and disk.  If the angular velocity of the core approaches the mass-shedding limit ($\Omega \approx \Omega_{\rm K}(R)$; $M = M_{\rm Kep}(J)$), the sign of the torque on the core must change, preventing it from gaining additional angular momentum.
We assume in this case that the core continues to accrete mass while remaining near the mass-shedding limit \citep{Popham&Narayan91,Bisnovatyi-Kogan93}.

The core's secular evolutionary track in $J-M$ space can then be simply integrated. Figure~\ref{fig:MJ} shows three examples of such tracks for different initial core masses and angular momenta $M_{\rm core}$, $J_{\rm core}$ (different colored solid curves, labeled A--C).  We model the core as a uniformly-rotating cold neutron star equilibrium solution obtained with the {\tt RNS} code \citep{Stergioulas+95} for a representative SLy EOS \citep{Douchin&Haensel01}.  By creating a grid of solutions in the $\{M$, $J\}$ solution space, we interpolate $j_{\rm Kep}$ within this parameter-space (taking $\zeta=1$) and numerically integrate $M(J)$ tracks, as shown.

Given that the equilibrium neutron star radius does not change too significantly as a function of $M$ and $J$, a useful analytic approximation can be obtained by assuming a constant $R_{\rm acc} = \zeta R$ in Eq.~(\ref{eq:j_Kep}).
This implies that accretion tracks follow
\begin{equation}
\label{eq:MofJ_accretion}
    M(J) = \left[ M_{\rm core}^{3/2} + \frac{3}{2 \sqrt{G \zeta R}} \left( J-J_{\rm core} \right), \right]^{2/3},
\end{equation}
in good agreement with the numerically-calculated curves in Fig.~\ref{fig:MJ}.

The fate of the merger-remnant is then directly determined by the core's evolutionary accretion track. The region where stable uniformly-rotating neutron star equilibrium solutions are permitted is bound between the two solid black curves in Fig.~\ref{fig:MJ}. The top curve shows the maximal mass of a uniformly-rotating neutron star $M_{\rm max}(J)$ (equal to $M_{\rm TOV}$ for non-rotating $J=0$ solutions). Above this curve, there is no equilibrium solution and the core undergoes gravitational collapse over dynamical timescales. The right boundary limits the core's rotation to below the mass-shedding limit, along which the equatorial angular velocity is equal to the Keplerian limit.
The two curves meet at a point that defines the maximum attainable mass for any solid-body rotating neutron star (see \citealt{Margalit+15} for further discussion of this point). This mass is typically a factor $\xi$ larger than the TOV mass, where
$\xi \simeq 1.2$ is insensitive to the EOS (e.g., \citealt{Baumgarte+00}).
A SMNS remnant is then typically defined as one where the total remnant mass is $M_{\rm TOV} < M_{\rm tot} < \xi M_{\rm TOV}$, while an indefinitely stable remnant requires $M_{\rm tot} < M_{\rm TOV}$.

As a consequence of the large orbital angular momentum of the original binary, BNS merger remnants are almost always born with total angular momenta exceeding the mass-shedding limit (to the right of the black curves in Fig.~\ref{fig:MJ}; e.g., \citealt{Radice+18}). Although a fraction of this initial angular momentum can be lost through various channels (e.g., GWs, neutrinos, post-merger outflows; see e.g. \citealt{Shibata+19}) on timescales comparable to the secular accretion timescale, $J_{\rm tot}$ is usually sufficiently large that it would be above (or near) the mass-shedding limit even accounting for such losses.
This has motivated the prevailing schematic view of post-merger remnants: consideration of the {\it total} post-merger remnant mass and angular momentum budget suggests that any SMNS or indefinitely-stable remnant ($M_{\rm tot} \lesssim 1.2 M_{\rm TOV}$) will redistribute its angular momentum over viscous timesecales to become a {\it dynamically-stable}, rapidly-rotating (at near breakup), equilibrium neutron star. The long-term evolution of this object is then governed by net angular momentum losses through, e.g. magnetic-dipole spindown, at fixed baryonic mass.  Such a trajectory is shown by the dotted portion of the yellow curve (case A).  However, as we now describe, our `two-zone' treatment of the post-merger remnant opens up the possibility for different outcomes, potentially challenging the traditional view discussed above.

\subsection{Evolutionary Tracks and Remnant Stability}

In the two-zone core+disk decomposition of the merger remnant, most of the total remnant mass $M_{\rm tot}$ is confined to a relatively slowly rotating core, while a small amount of centrifugally-supported mass (the disk) carries most of the total angular momentum $J_{\rm tot}$.  The remnant's evolution and stability should therefore be assessed based on properties of the core, which differ substantially from the total remnant properties (compare the stars to the triangle in Fig.~\ref{fig:MJ}).  The evolution of the core over accretion timescales depends sensitively on its initial properties.  Figure~\ref{fig:MJ} shows three illustrative cases with different initial values of $\{J_{\rm core}$, $M_{\rm core}\}$.
In case A with a relatively low initial mass $M_{\rm core} \approx 0.78 M_{\rm tot}$ (yellow) the core accretes mass, roughly following Eq.~(\ref{eq:MofJ_accretion}), until it hits the mass-shedding curve. Subsequently, the core accretes by rising up this mass-shedding curve until the disk mass has been depleted and $M \simeq M_{\rm tot}$ (for simplicity, we neglect mass-loss due to disk winds).  This marks the end of the remnant's viscous evolution phase (yellow circle).  Over longer timescales $t \gg t_{\rm visc, disk}$ the remnant then spins down through non-accretion processes (e.g., via magnetic-dipole spindown), along a track of constant baryonic mass and decreasing $J$ (dotted yellow curve). In effect, case~A behaves like the typically-envisioned SMNS-remnant evolution described in the previous paragraph.  The outcome would be a remnant that survives over long timescales and that loses a large amount rotational energy $\gtrsim 10^{52}-10^{53}$ erg into the merger environment. This likely has an observable impact on the electromagnetic counterparts of such mergers (e.g., \citealt{Metzger&Piro14}).

If by contrast the initial core mass is larger and/or the initial core angular momentum is smaller, then the outcome can change qualitatively. The red star (case~C) in Fig.~\ref{fig:MJ} illustrates an example of this kind.  In this case the secular accretion track increases the core mass such that it first hits the $M_{\rm max}(J)$ limit instead of the mass-shedding curve.  This implies that the remnant will undergo an accretion-induced collapse (AIC) and hence never pass through a spindown phase, thereby substantially reducing the impact of the remnant on the electromagnetic emission from the merger.  This is in stark contrast to the traditional view of BNS merger remnants with masses within the (traditionally-defined) SMNS window.  Finally, case B (blue curve) shows an intermediate case where the core does spin-up through accretion and then undergoes a long-term spindown phase (dotted-blue portion of the curve); however, because the core does not accrete to reach break-up rotation, it loses less angular momentum and rotational energy than in case A.

Within our idealized two-zone formalism, there exists a critical curve $M_{\rm AIC}(J)$ that delineates between the accretion-induced and spin-down-induced collapse regimes (dot-dashed black curve in Fig.~\ref{fig:MJ}). Scenarios similar to case C where the remnant collapses without going through a spindown phase have core masses that fall above this critical curve, whereas case A/B-like evolutionary tracks occur if the core mass is $M_{\rm core} < M_{\rm AIC}(J_{\rm core})$.  An analytic estimate of this condition can be obtained from the intersection of Eq.~(\ref{eq:MofJ_accretion}) with the maximal-mass curve $M_{\rm max}(J).$  The latter can be roughly approximated with the following implicit relation for $M_{\rm max}$,
\begin{eqnarray}
\label{eq:Mmax_of_J}
    M_{\rm max}(J) 
    &\approx& M_{\rm TOV} \left[ 1 + (\xi-1) \left(\frac{\Omega}{\Omega_{\rm K}}\right)^2 \right] \nonumber \\
    &=& M_{\rm TOV} \left[ 1 + \frac{(\xi-1) J^2}{\eta^2 G M_{\rm max}^3 R} \right] .
\end{eqnarray}
This functional form is motivated by the $\propto J^2$ scaling of rotational-support in the Newtonian limit and it ensures that $M_{\rm max}(J_{\rm max}) = \xi M_{\rm TOV}$, per our definition of $\xi$.  In the second equality in Eq.~(\ref{eq:Mmax_of_J}) we defined the dimensionless
moment of inertia
\begin{equation}
    \eta 
    \equiv I / M R^2
\end{equation}
where $I \sim 10^{45} \, {\rm g \, cm}^2$ is the NS moment of inertia.
For the SLy EOS used in the example calculations above, $\eta$ spans a narrow range $\eta \simeq 0.3-0.4$ throughout the surveyed $M$-$J$ parameter space.
More accurate quasi-universal relations have been found for $M_{\rm max}(J)$ and $\eta$ as a function of neutron star compactness (e.g., \citealt{Breu&Rezzolla16}) and could be employed here instead, however our above assumptions (Eq.~\ref{eq:Mmax_of_J} and constant $\eta$) are sufficient given other uncertainties associated with our toy-model.

The critical curve $M_{\rm AIC}(J)$ is defined such that a core whose initial mass and angular momentum fall on this curve would reach the maximal-mass curve exactly at the termination of the accretion phase, i.e. such that $M(J) = M_{\rm max}(J) = M_{\rm tot}$. Using Eqs.~(\ref{eq:MofJ_accretion},\ref{eq:Mmax_of_J}) and the above condition, we find that
\begin{eqnarray}
\label{eq:M_AIC}
&&    M_{\rm AIC}(J;M_{\rm tot}) 
    \approx \\ \nonumber
    && M_{\rm tot} \left[ 1 - \frac{3}{2}\eta \sqrt{\frac{M_{\rm tot}/M_{\rm TOV}-1}{\zeta (\xi-1)}} + \frac{3J}{2\sqrt{G M_{\rm tot}^3 \zeta R}} \right]^{2/3}
\end{eqnarray}
where we recall that $\xi \simeq 1.2$ and $\eta \approx 0.4$.
The condition for a remnant to undergo accretion-induced collapse during its viscous evolution is therefore,
\begin{equation}
\label{eq:AIC_condition}
    {\rm AIC:} ~~~ M_{\rm core} \geq M_{\rm AIC}(J_{\rm core}) 
    .
\end{equation}
If the core does not accrete the total remnant mass $M_{\rm tot}$ (e.g., if some mass is lost over viscous timescales due to disk winds) then $M_{\rm tot}$ in Eq.~(\ref{eq:M_AIC}) should be replaced by the appropriate final mass.\footnote{In deriving Eq.~(\ref{eq:M_AIC}) we have implicitly assumed that the gravitational mass of the core in its final state is equal to the total remanant's initial gravitational mass. It would be more appropriate to apply this condition to the {\it baryonic} mass of the core and remnant; however the error introduced by our simplifying approach is much smaller than other sources of uncertainty in this derivation (note that contours of constant baryonic mass [dotted curves] in Fig.~\ref{fig:MJ} are nearly horizontal).}

\begin{figure*}
    \centering
    \includegraphics[width=0.85\textwidth]{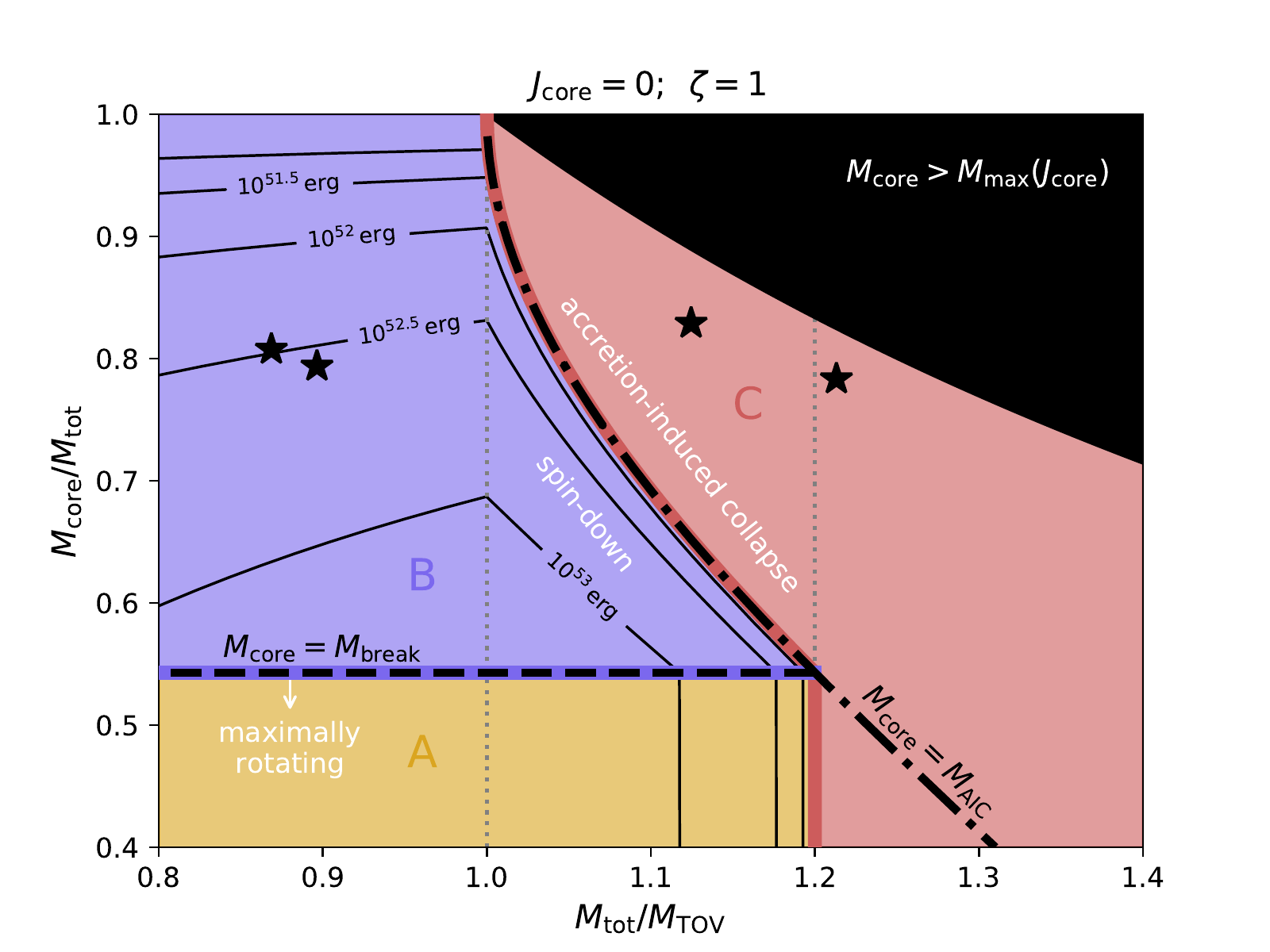}
    \includegraphics[width=0.45\textwidth]{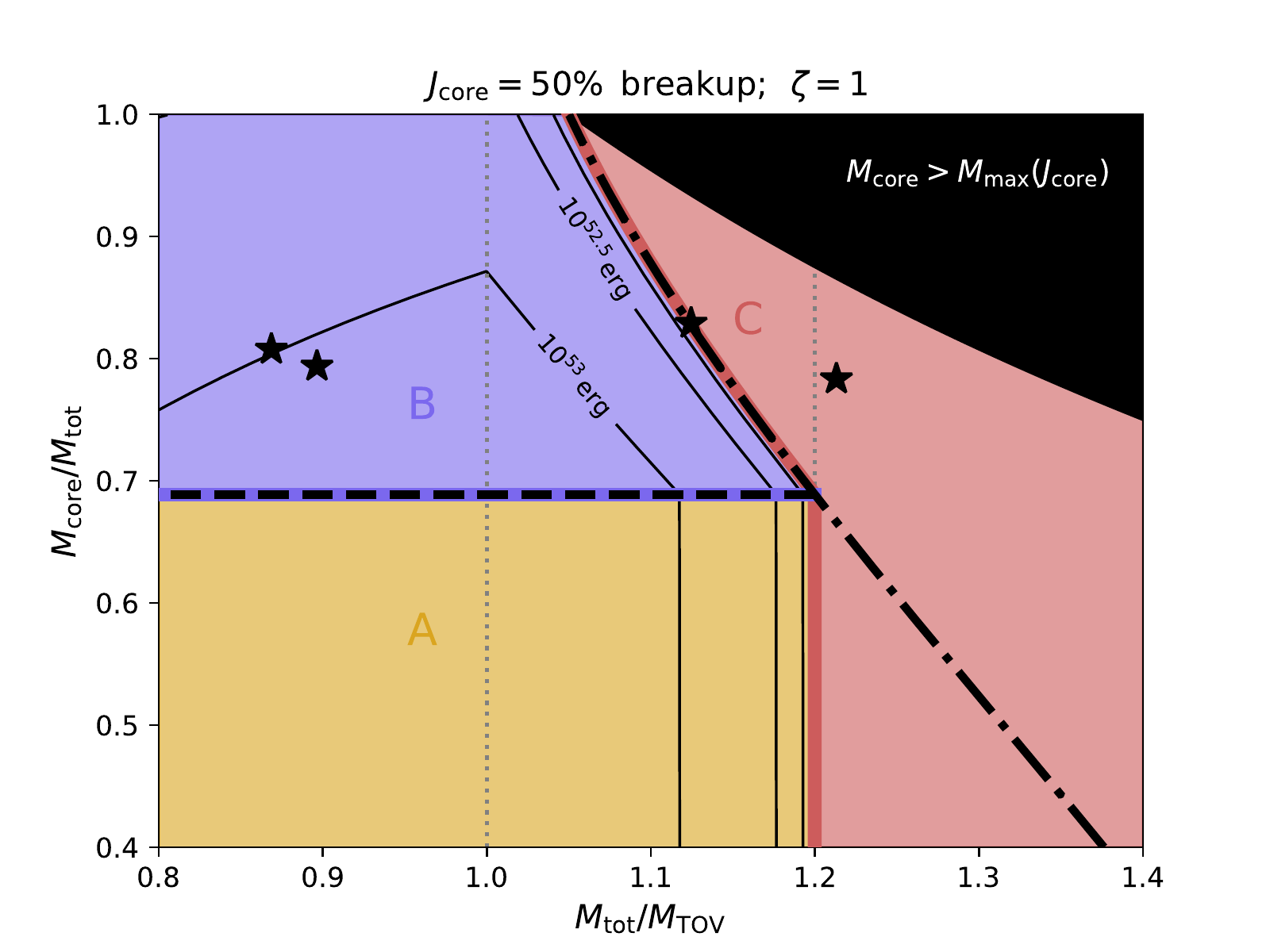}
    \includegraphics[width=0.45\textwidth]{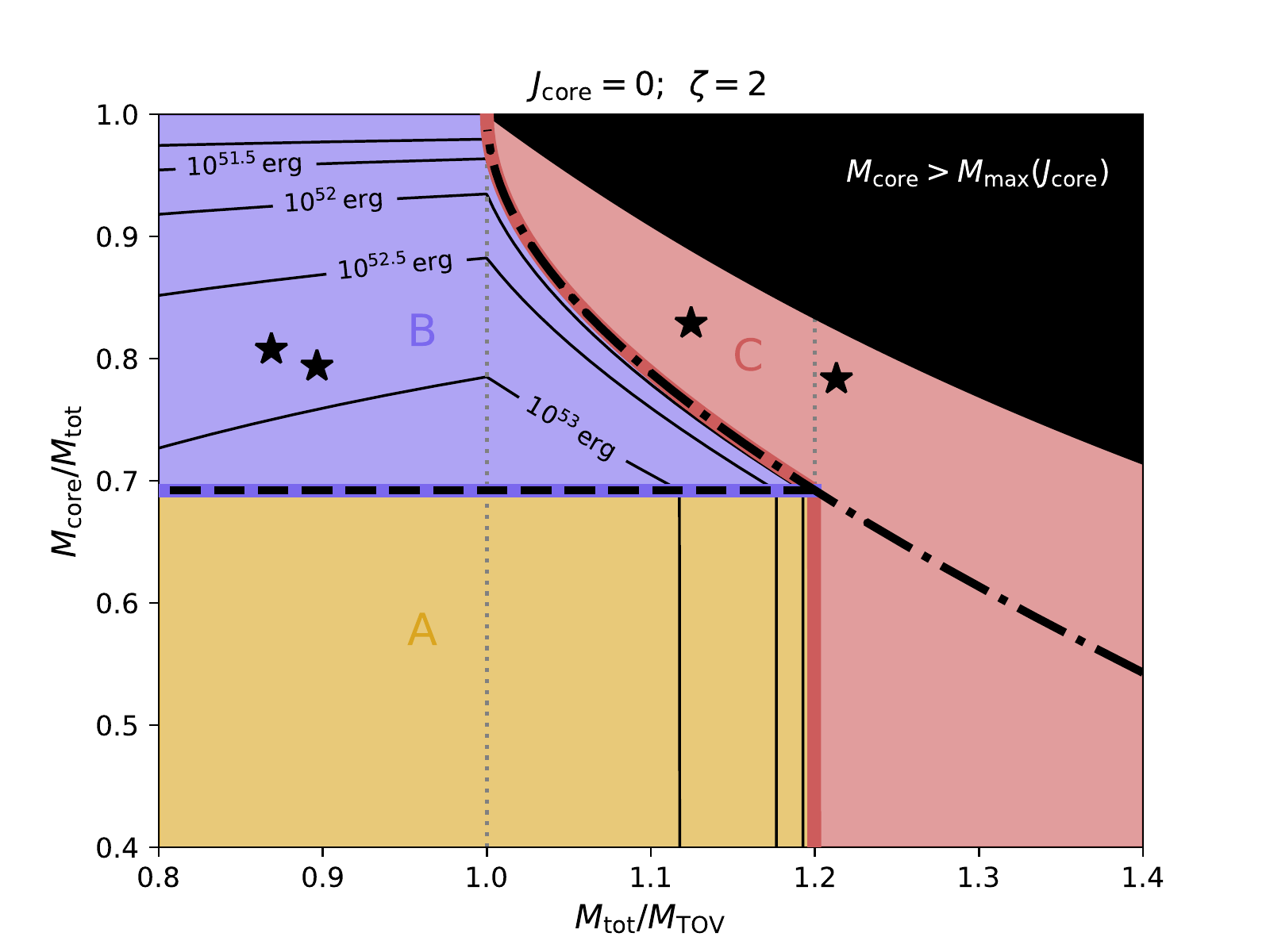}
    \caption{
    Merger remnant outcomes as a function of the total remnant mass $M_{\rm tot}$ normalized to the TOV mass and the fractional initial mass $M_{\rm core}$ in the comparatively slowly rotating core.  Each panel corresponds to different assumptions about the initial core angular momentum (e.g., $J_{\rm core} = 0$ and $J_{\rm core} = 50\%$ breakup) and the effective lever arm for adding specific angular momentum to the core ($\zeta \equiv R_{\rm acc}/R = \{1,2\}$; Eq.~\ref{eq:j_Kep}) as marked.  The three qualitatively different outcomes illustrated by the tracks A, B, C in Fig.~\ref{fig:MJ} are denoted by yellow, blue, and red regions.  Remnants formed from massive mergers (large $M_{\rm tot}/M_{\rm TOV}$) with a large fraction of their mass in a slowly rotating core (large $M_{\rm core}/M_{\rm tot}$) collapse as a result of accretion from the envelope/disk (Region C; `accretion-induced collapse').  By contrast, lower mass remnants and/or those with smaller cores remain as stable neutron stars even after the entire envelope has been accreted (`spindown'), whether the core is at that point rotating at breakup (Region A) or sub-critically (Region B). Solid black contours indicate the `extractable' rotational energy that can be removed from the remnant into the environment, $T_{\rm ext}$ (e.g., via magnetic spin-down; Eq.~\ref{eq:T_ext}).  The black region in the upper right corner is forbidden because cores of mass $> M_{\rm max}(J_{\rm core})$ are dynamically unstable.  Black stars denote estimates of $\{M_{\rm core},M_{\rm tot}\}$ based on individual numerical relativity BNS merger simulations for several different EOS (SHT, APR4, MS1, and SFHo; from \citealt{Kastaun+16,Ciolfi+17,Kastaun&Ohme21}, respectively). The calculations in this figure assume $\xi=1.2$, $\eta=0.4$, $M_{\rm TOV} = 2.1 M_\odot$, and $R = 12 \, {\rm km}$.
    }
    \label{fig:Mcore_Mtot}
\end{figure*}

A similar estimate can be made for the curve $M_{\rm break}(J)$ that distinguishes between cores that hit the mass-shedding curve and are rotating at break-up following their viscous evolution (similar to case A) versus cores whose maximal rotation rate is lower (case B). This curve is shown as the dashed-black line in Fig.~\ref{fig:MJ}. An analytic estimate of it can be obtained by equating Eq.~(\ref{eq:MofJ_accretion}) to the mass-shedding limit, $M_{\rm Kep}(J)$. The latter is roughly given by setting the star's rotation rate $\Omega = J/I$ to the Keplerian limit, which yields
\begin{equation}
\label{eq:M_Kep}
    M_{\rm Kep}(J) \approx \left( J^2 / \eta^2 G R \right)^{1/3} .
\end{equation}
Combined with Eq.~(\ref{eq:MofJ_accretion}), we find that merger remnants will be spinning at break-up at the end of the accretion phase if the initial core mass is below
\begin{equation}
\label{eq:M_break}
    M_{\rm break}(J;M_{\rm tot}) \approx M_{\rm tot} \left[ 1 - \frac{3 \eta}{2 \zeta^{1/2}} + \frac{3J}{2\sqrt{G M_{\rm tot}^3 \zeta R}} \right]^{2/3} .
\end{equation}

A special use-case of Eq.~(\ref{eq:M_AIC}) occurs if the total remnant mass falls within the hypermassive region, i.e. $M_{\rm tot} > \xi M_{\rm TOV}$. In this case the two curves $M_{\rm AIC}(J)$ and $M_{\rm break}(J)$ merge into one, stemming from the `tip' of the SMNS-solution triangle where the maximal-mass $M_{\rm max}$ and mass-shedding $M_{\rm Kep}$ curves meet. This is equivalent to the substitution $M_{\rm tot} = \xi M_{\rm TOV}$ in Eq.~(\ref{eq:M_AIC}), which simplifies the expression. In particular, for the limiting case where the initial core angular momentum is $J_{\rm core} \approx 0$, the condition for the hypermassive remnant to collapse 
before the core reaches breakup rotation is that
$M_{\rm core} > \xi \left( 1 - 3\eta/2 \right)^{2/3} \approx 0.7 M_{\rm TOV}$.  By contrast, if $M_{\rm core} \lesssim 0.7M_{\rm TOV}$, then a hypermassive remnant will collapse by ascending the mass-shedding line to the tip of the SMNS-solution triangle.  These different outcomes would have implications for the quantity of mass remaining in an accretion disk surrounding the newly-formed black hole, being (perhaps counter-intuitively) larger in cases when the initial disk-mass $M_{\rm d,0} \simeq M_{\rm tot}-M_{\rm core}$ is smaller for systems of the same total mass $M_{\rm tot}$.

Finally, we estimate the amount of rotational energy $T_{\rm ext}$ that is extractable via spindown (see \citealt{Metzger+15} for the introduction of this concept). Given our assumption of solid-body rotation, the rotational energy in the Newtonian limit is simply $T = J^2/2\eta MR^2$. 
The maximum rotational energy is determined by the angular momentum at the mass shedding limit, where $M_{\rm Kep}(J) = M_{\rm tot}$ (Eq.~\ref{eq:M_Kep}). This gives $T = \eta G M_{\rm tot}^2 / 2R$. A merger remnant whose core mass is $M_{\rm core} > M_{\rm break}$ (above the dashed curve in Fig.~\ref{fig:MJ}) will begin the spindown phase with rotational energy that is a fraction $f_{\rm max}$ smaller than this maximal value. Furthermore, the extractable rotational energy can be lower than the initial energy if the remnant undergoes spindown-induced collapse. In this case a fraction $f_{\rm min}$ of the maximal rotational energy is ``lost'' when the remnant collapses. 
The extractable rotational energy can therefore be estimated as
\begin{equation}
\label{eq:T_ext}
    T_{\rm ext} 
    \approx 
    \begin{cases}
    0 ~,~ M_{\rm core} \geq M_{\rm AIC} ~\text{or}~ M_{\rm tot} \geq \xi M_{\rm TOV}
    \\
    \dfrac{\eta G M_{\rm tot}^2}{2R}  \left( f_{\rm max} - f_{\rm min} \right) ~,~ {\rm else}
    \end{cases}
\end{equation}
where
\begin{align}
\label{eq:f_max}
    &f_{\rm max} 
    = 
    \frac{4\zeta}{9\eta^2} 
    \left[ 1 - \left(\frac{M_{\rm core}}{M_{\rm tot}}\right)^{3/2} \right. &
    \\ \nonumber
    &~~~~~~~~~~~~~~~~~~~~\left. + \frac{3 J_{\rm core}}{2 \sqrt{G M_{\rm tot}^3 \zeta R}} \right]^2
    &;~ M_{\rm core} > M_{\rm break}
    \\ \nonumber
    &f_{\rm max} = 1 &;~ M_{\rm core} \leq M_{\rm break}
\end{align}
and
\begin{equation}
    f_{\rm min} = 
    \begin{cases}
    \dfrac{M_{\rm tot}/M_{\rm TOV} - 1}{\xi - 1} &, M_{\rm tot} > M_{\rm TOV}
    \\
    0 &, M_{\rm tot} \leq M_{\rm TOV}
    \end{cases}
    .
\end{equation}


\subsection{Parameter Space of Merger Outcomes}

With the analytic estimates above, we can now survey the parameter space of merger outcomes implied by our two-zone toy model.
As discussed in the previous subsections and illustrated in Fig.~\ref{fig:MJ}, the merger outcome depends on properties of the binary and merger---specifically the total binary mass $M_{\rm tot}$ and post-merger core mass and angular momentum $M_{\rm core}$, $J_{\rm core}$---as well as properties of the EOS and  accretion physics (in our simple model encapsulated via $M_{\rm TOV}$, $R$, $\xi$, and $\zeta$).

Figure~\ref{fig:Mcore_Mtot} shows this phase space in terms of the TOV-normalized binary mass $M_{\rm tot}/M_{\rm TOV}$ and the post-merger core mass-fraction $M_{\rm core}/M_{\rm tot}$. 
The red shaded area shows regions where the remnant undergoes accretion-induced collapse, similar to the red track (case C) illustrated in Fig.~\ref{fig:MJ}. Blue and yellow regions instead delineate the parameter space where the remnant goes through a spindown phase prior to reaching its final state: collapse (if $M_{\rm tot}>M_{\rm TOV}$) or an indefinitely stable neutron star ($M_{\rm tot}<M_{\rm TOV}$).
Within the yellow region the core starts this spindown phase maximally rotating (at the mass-shedding limit), similar to track A in Fig.~\ref{fig:MJ}. This implies $f_{\rm max}=1$ and a large amount of extractable rotational energy (black contours in Fig.~\ref{fig:Mcore_Mtot}; Eq.~\ref{eq:T_ext}). Case-B-like tracks where the core never achieves maximal rotation ($f_{\rm max} < 1$) are enclosed within the blue region.
The boundary between the yellow and blue regions is set by the condition $M_{\rm core} = M_{\rm break}(J_{\rm core},M_{\rm tot})$ (dashed curves in Figs.~\ref{fig:MJ},\ref{fig:Mcore_Mtot}; Eq.~\ref{eq:M_break}). Similarly, the boundary between spindown- and accretion-induced collapse (blue and red regions) is determined by the condition $M_{\rm core} = M_{\rm AIC}(J_{\rm core},M_{\rm tot})$ (dot-dashed curves in Figs.~\ref{fig:MJ},\ref{fig:Mcore_Mtot}; Eq.~\ref{eq:M_AIC}).
Our toy-model breaks down in the black shaded region where $M_{\rm core} > M_{\rm max}(J_{\rm core})$ and no stable solid-body equilibrium core solution is possible. Naively, mergers within this parameter-space may be thought to undergo prompt-collapse. 
However this naive condition need not correspond to more detailed estimates of the prompt-collapse threshold (e.g., \citealt{Bauswein+13,Bauswein&Stergioulas17}) given that 
prompt-collapse occurs within dynamical timescales after merger, when an equilibrium core is not well-defined.

Figure~\ref{fig:Mcore_Mtot} illustrates that mergers with larger $M_{\rm tot}$ and/or $M_{\rm core}$ generally result in less stable remnants that can collapse more rapidly.
Within the standard view, remnants with $1 < M_{\rm tot}/M_{\rm TOV} \lesssim 1.2$ are classified as SMNSs and are assumed to be near maximally-rotating and undergo spindown-induced collapse. Under this paradigm, the entire region between $M_{\rm tot}/M_{\rm TOV} = 1$ and $M_{\rm tot}/M_{\rm TOV} = \xi \simeq 1.2$ (vertical-dotted curves in Fig.~\ref{fig:Mcore_Mtot}) should be painted yellow. Here, we have argued that the core mass plays a critical role in altering this picture. In particular, Fig.~\ref{fig:Mcore_Mtot} shows that a diverse set of outcomes are possible within the traditionally-classified SMNS regime, dependant on the initial core mass.

The importance of the core mass within this new paradigm motivates more detailed investigation of this property. Black stars in Fig.~\ref{fig:Mcore_Mtot} show the core mass-fraction estimated from a handful of numerical-relativity simulations \citep{Kastaun+16,Ciolfi+17,Kastaun&Ohme21}. Specifically, these points show the results (reported in baryonic masses) of equal mass-ratio BNS merger simulations with similar binary masses and different EOS:
SHT ($M_{\rm tot}^{\rm b} = 3.03 M_\odot$, $M_{\rm TOV}^{\rm b} = 3.38 M_\odot$, and `bulk mass' $M_{\rm core}^{\rm b} = 2.405 M_\odot$; \citealt{Kastaun+16}); 
APR4 ($M_{\rm tot}^{\rm b} = 2.98 M_\odot$, $M_{\rm TOV}^{\rm b} = 2.65 M_\odot$, and bulk mass $M_{\rm core}^{\rm b} = 2.47 M_\odot$; \citealt{Ciolfi+17}); 
MS1 ($M_{\rm tot}^{\rm b} = 2.91 M_\odot$, $M_{\rm TOV}^{\rm b} = 3.35 M_\odot$, and bulk mass $M_{\rm core}^{\rm b} = 2.35 M_\odot$; \citealt{Ciolfi+17}); 
and SFHo ($M_{\rm tot}^{\rm b} = 3 M_\odot$, $M_{\rm TOV}^{\rm b} = 2.47 M_\odot$, and `core-equivalent' mass $M_{\rm core}^{\rm b} = 2.35 M_\odot$ estimated from Fig.~8 of \citealt{Kastaun&Ohme21}).\footnote{
We omit results reported in \cite{Ciolfi+17} for the H4 EOS because the bulk mass in this case is more ambiguous, showing substantial evolution during the $\sim 20\,{\rm ms}$ following merger and till collapse (see Fig.~14 of that work).}
These results suggest that the core mass-fraction is $M_{\rm core}/M_{\rm tot} \sim 0.8$ for a range of $M_{\rm TOV}$. 
A roughly constant core mass-fraction $M_{\rm core}/M_{\rm tot} \sim 0.8$ is  also consistent with the results of Newtonian smoothed-particle hydrodynamic NS merger simulations reported in \cite{Fryer+15}. 
Further exploration is however needed to verify the robustness of these results and identify potential trends with binary parameters or EOS properties.
Furthermore, the initial core angular momentum is typically not reported in published simulations.
The angular-velocity profile of BNS merger remnants (e.g., Fig.~\ref{fig:schematic}) suggests that $J_{\rm core}$ is low, however more quantitative estimates would be useful in the context of the two-zone model presented in this paper.
As a rough approximation, we use the central rotation rates quoted in \cite{Ciolfi+17} to estimate that $J_{\rm core} \sim 30\%$ ($15\%$) of breakup for the equal-mass APR4 (MS1) simulations.
This falls within the range shown in Fig.~\ref{fig:Mcore_Mtot}.\footnote{\cite{Fryer+15} quote $J_{\rm core}$ values that imply a broad range extending to large angular momenta, $J_{\rm core} > 50\%$ breakup. We note however that: (a) the definition of ``core'' in that work may not perfectly coincide with our current definition; (b) these early simulations employed only Newtonian gravity.}

\subsection{Implications on the NS EOS}

We conclude with a brief discussion of the implications of our model for the ability to constrain the EOS-dependent TOV mass $M_{\rm TOV}$ from BNS merger events. Such constraints have been derived using combined information on the total mass of the binary (and hence---once accounting for various sources of ejecta---of the remnant, $M_{\rm tot}$) based on the gravitational wave inspiral signal with electromagnetic observations which constrain the nature of the final merger product (see e.g. \citealt{Margalit&Metzger19}).  As a proxy for the latter, we focus on the extractable rotational energy, $T_{\rm ext}$, which grows rapidly entering the SMNS regime.  In the traditional view (e.g., \citealt{Lawrence+15}) if a SMNS can be excluded observationally (e.g., by detection of a gamma-ray burst or by placing an upper limit on $T_{\rm ext} \ll 10^{52}$ erg), then one can place an upper limit $M_{\rm TOV} \lesssim M_{\rm tot}/\xi$, or slightly more precisely---by accounting for conversion between baryonic and gravitational mass---the limit is roughly 
\begin{equation}
\label{eq:M_TOV}
    M_{\rm TOV} \lesssim \frac{\sqrt{1+0.3 M_{\rm tot}^{\rm b} / \xi}-1}{0.15} ,
\end{equation}
where $M_{\rm tot}^{\rm b}$ is the total baryonic mass of the remnant \citep{Margalit&Metzger17}.
For GW170817 ($M_{\rm tot}^{\rm b} \lesssim 3.06 M_\odot$) this resulted in $M_{\rm TOV} \lesssim 2.2 M_{\odot}$ (e.g., \citealt{Margalit&Metzger17,Shibata+17,Rezzolla+18}; see also \citealt{Lawrence+15}).  By contrast, applying the same criterion using the new $T_{\rm ext}$ contours in Fig.~\ref{fig:Mcore_Mtot} for values of $M_{\rm core}/M_{\rm tot} \sim 0.8$ estimated from BNS merger simulations (stars in Fig.~\ref{fig:Mcore_Mtot}; e.g., \citealt{Kastaun+16,Ciolfi+17,Kastaun&Ohme21}), we find $M_{\rm TOV} \lesssim 2.5 M_{\odot}$ and $\lesssim 2.3 M_{\odot}$ in the cases of $J_{\rm core} = 0$ and $J_{\rm core} = 50\%$ breakup, respectively.
This can be estimated directly from Eq.~(\ref{eq:M_AIC}) by defining
\begin{equation}
    \xi^\prime \equiv 1 + (\xi-1)
    f_{\rm max}
    \label{eq:xiprime}
\end{equation}
such that 
mergers with $M_{\rm tot} \geq \xi^\prime M_{\rm TOV}$ 
result in accretion-induced collapse and do not pass through a spindown phase (Eq.~\ref{eq:AIC_condition}).
Here $f_{\rm max}$ is a function of $M_{\rm core}$, $J_{\rm core}$ defined in Eq.~(\ref{eq:f_max}).
Within our new picture, a revised TOV constraint can be obtained by using Eq.~(\ref{eq:M_TOV}) with $\xi \to \xi^\prime$. 
It is always the case that $\xi^\prime \leq \xi$, and therefore the TOV constraint is weakened.

Finally, we mention several deficiencies of our model, which are likely to change the quantitative conclusions and should be explored in future work.  Although we include general relativistic effects in calculating the stability of the core, we have used Newtonian gravity or neglected strong-field effects in several places throughout the analysis (e.g., not carefully distinguishing between gravitational and baryonic mass).  Our analysis also neglects the effects of thermal pressure on the stability of the SMNS \citep{Kaplan+14}, which can be important on timescales of hundreds of milliseconds to seconds post merger. Thermal pressure in the outer layers of the star generally acts to reduce the maximum mass of the SMNS remnant by up to 8\% (mainly by reducing the angular velocity at the mass-shedding limit).  We also do not consider the effects of phase transitions on the NS stability (e.g., \citealt{Keil&Janka95}). 
Finally, we have not accounted for non-axisymmetric instabilities that can take place in rapidly rotating stars prior to reaching the mass shedding limit and which can cause an outward angular momentum flux \citep{Takiwaki+21,Pan+21}.

\section{Discussion and Conclusions}
\label{sec:conclusions}

We have explored the efficacy of angular momentum transport inside very rapidly rotating ($\sim{\rm ms}$ spin-period) PNS formed in core collapse supernovae and BNS mergers by means of analytic estimates, evaluated by imposing an angular velocity profile by hand on top of a one-dimensional PNS cooling evolution model.  A drawback of our work is that the PNS model we employ as the background state does not self-consistently include the (likely substantial) effects of rotation.  Nevertheless, it is still sufficient to yield new qualitative insights, given the order-of-magnitude aspiration of our estimates and the many open theoretical questions in the efficiency of angular momentum transport (e.g., effective $\alpha$-viscosity parameter arising from MRI, or the saturation strength of the Spruit-Tayler dynamo).  Some of these processes, while extensively considered in the stellar evolution literature, have to our knowledge not been explored in the context of BNS merger remnants.

At essentially all radii and times of interest in both core collapse and BNS merger cases, the timescale for angular momentum transport is short compared to the cooling time (time since core bounce or merger) over which the PNS thermal properties are evolving or over which mass is accreted from an external Keplerian disk (Figs.~\ref{fig:time_radius_cc}, \ref{fig:time_radius_remnant}).  
In regions where $d\Omega(r)/dr > 0$, the combination of the TS dynamo and convection operate efficiently (in stably and unstable-stably stratified regions, respectively).  At the largest radii where $d\Omega(r)/dr < 0$, the MRI typically dominates.  
We have also derived a revised analytic estimate of neutrino viscosity that is applicable to arbitrary neutrino degeneracy (Eqs.~\ref{eq:nunu_full},\ref{eq:nunu_cases}; see Appendix~\ref{sec:Appendix}). Even though neutrino viscosity is significantly enhanced due to $\nu_e$ degeneracy in central regions of the star, we find that neutrino viscosity is unimportant in the configurations we have investigated (in the sense that $t_\nu \gg$ other timescales of interest).

One uncertainty in our analysis is that it is not clear whether convection efficiently enforces solid body rotation.   This is not the case, e.g., in simulations of slowly rotating red-giant convection (e.g., \citealt{Brun2009}; see also \citealt{Kissin_2015} for a theoretical analysis of this problem).  The numerical evidence does suggest, however, that an approach to solid body rotation is more likely in the rapidly rotating limit considered here  (e.g.,~\citealt{GASTINE2012428,YADAV2013185,Mabuchi_2015}).   In addition, the limited radial extent of PNS convection zones likely  mitigates the effects of moderate residual differential rotation on the PNS stability.

A goal of future numerical work should be to include the physical sources of viscosity introduced in this work into self-consistent numerical relativity simulations of the post-merger remnant evolution.    Our  analysis suggests that angular momentum transport due to a combination of mechanisms is rapid throughout most of the neutron star.   This suggests that a fruitful first path is to explore the long-term secular evolution of rotating PNSs and merger remnants using sub-grid `viscous' models motivated by our analytic considerations.  Such sub-grid models can be refined using local simulations of proto-NS convection and magnetic instabilities.    A more ambitious goal is to include physical sources of viscosity directly in global magneto-neutrino-hydrodynamical simulations.   Neutrino viscosity should in principle be straightforward to include given the neutrino properties already available in simulations which include neutrino transport (in a manner similar to the implementation of an $\alpha$-viscosity in \citealt{Fujibayashi+20}).   The growth and saturation of the MRI can likewise potentially be directly resolved inside the remnant (e.g., \citealt{Kiuchi+18}) as can convective instabilities (which are significantly more vigorous in proto-NSs than in stellar interiors).  Resolving the dynamo processes responsible for setting the saturation strength of the Spruit-Tayler or other small-scale magnetic dynamos is likely to be the most challenging numerically (e.g., \citealt{Skoutnev+21}). 
    
Our results could have implications for both core collapse supernovae and BNS mergers.  Insofar that the dissipation from various forms of effective viscosity transform the energy in differential rotation into heat, and that the free energy available in a PNS rotating near break-up is non-negligible compared to its internal thermal energy during the Kelvin-Helmholtz phase, our finding of rapid angular momentum transport throughout optically-thick regions of the star could have a quantitative impact on the cooling evolution, and hence neutrino emission properties, of rapidly spinning PNS (e.g., with implications for the supernova explosion process; \citealt{Thompson+05}).

 Our finding that angular momentum transport should be rapid in the core of BNS merger remnants, implies that the core will quickly evolve to a rigidly rotating state, consistent with the evolution found in numerical relativity simulations which impose an $\alpha$-viscosity everywhere by hand (e.g., \citealt{Fujibayashi+20}).  Motivated by this structure, we develop a model which follows the growth and evolution of the core via accretion of mass and angular momentum from the surrounding quasi-Keplerian envelope/disk.  We use this model to elucidate the remnant's stability and ultimate fate in terms of its initial properties and unknown features of the neutron star EOS, particularly the TOV mass.
    
Depending on the initial mass and angular momentum of the remnant core $\{M_{\rm core}, J_{\rm core}\}$, we identify three qualitatively distinct evolutionary paths (Figs.~\ref{fig:MJ}, \ref{fig:Mcore_Mtot}).  For low $M_{\rm core}$/high $J_{\rm core}$ (Region A), the core reaches the mass-shedding limit and accretes along this line.  For remnants with a total masses obeying the nominal SMNS limit $M_{\rm tot} \lesssim \xi M_{\rm TOV} \approx 1.2M_{\rm TOV}$, the end product is a nearly maximally-spinning SMNS; such an object can only collapse to a black hole via subsequent spin-down (e.g., due to magnetic dipole braking), accompanied by the transfer of significant rotational energy into the surroundings (Eq.~\ref{eq:T_ext}).  For mergers generating cores with somewhat higher $M_{\rm core}$/lower $J_{\rm core}$ (Region B), the star will not reach maximal rotation but nevertheless will still survive as a SMNS (which however requires less rotational energy loss to collapse).  
    
By contrast, in the case of highest $M_{\rm core}$/lowest $J_{\rm core}$ (Region C; as may be suggested by numerical relativity BNS merger simulations) the core will reach the collapse line ($M_{\rm max}(J)$ in Fig.~\ref{fig:MJ}) before accreting the entire envelope.  In this case, a black hole will form relatively promptly (likely within the first seconds or less after the merger) and the rotational energy available to be injected into the environment by the remnant will be drastically reduced.  The potential for an evolutionary path which results in accretion-induced black hole formation, even for a total system mass and angular momentum which permits SMNS formation, runs counter to the basic assumption of previous work (e.g., \citealt{Margalit&Metzger17,Shibata+17,Rezzolla+18,Shibata+19}).  A reduced fraction of BNS mergers which generate long-lived stable magnetar remnants is consistent with the lack of evidence for such large energy injection based on observations of short gamma-ray bursts (e.g., \citealt{Metzger&Bower14,Horesh+16,Schroeder+20,Beniamini&Lu21}).
    
If confirmed by future work, our finding that the fate of the merger remnant depends on the initial core properties $\{M_{\rm core},J_{\rm core}\}$ in addition to the total binary mass $M_{\rm tot}/M_{\rm TOV}$ would also have key implications for predicting the range of diversity in the electromagnetic counterparts of BNS mergers and the ability of using an observationally-inferred merger outcomes to constrain the neutron star EOS (e.g., \citealt{Margalit&Metzger19}).  In particular, even if a robust constraint can be placed on the merger lifetime or the remnant rotational energy from a merger event with $M_{\rm tot}$ measured from gravitational wave observations, the upper limit one can place on $M_{\rm TOV}$ will be weakened for Region C-like evolution (Eq.~\ref{eq:xiprime}), e.g. by up to a few tenths of a solar mass in the case of GW170817.  While a limited set of numerical relativity simulations seem to tentatively support merger remnants undergoing C-type evolution (stars in Fig.~\ref{fig:Mcore_Mtot}), the qualitatively new picture outlined in this work motivates further numerical simulation work to systematically quantify the post-merger core properties $\{ M_{\rm core},J_{\rm core} \}$ for different EOS and initial binary masses and mass-ratios.

\begin{acknowledgements}
We thank Todd Thompson for helpful comments on the manuscript and for 
correspondence regarding neutrino viscosity calculations.
We additionally thank Daniel Siegel and Wolfgang Kastaun for helpful discussions and comments.
B.M. is supported by NASA through the NASA Hubble Fellowship grant \#HST-HF2-51412.001-A awarded by the Space Telescope Science Institute, which is operated by the Association of Universities for Research in Astronomy, Inc., for NASA, under contract NAS5-26555.
B.D.M. acknowledges support from the National Science Foundation (grant number AST-2002577).
The Flatiron Institute is supported by the Simons Foundation.
E.Q. is supported in part by a Simons Investigator Award from the Simons Foundation.
This work has been assigned a document release number LA-UR-22-25436.
\end{acknowledgements}

\appendix
\section{Neutrino Viscosity with Degenerate Neutrinos}
\label{sec:Appendix}

Neutrino viscosity in PNS and BNS merger remnants is dominated by scattering of electron neutrinos. Previous approximate analytic formulae for $\nu_\nu$ assume that neutrinos are non-degenerate (e.g. \citealt{Keil+96,Thompson&Duncan93}). However, deep within the neutron star core the electron neutrino chemical potential $\mu_{\nu_e}$ can be significant and the neutrino degeneracy parameter $\eta_\nu \equiv \mu_\nu / k T$ large. 
Neutrino degeneracy affects the scattering cross sections, the neutrino energy density, and Pauli-blocking factors, resulting in an enhanced viscosity (\citealt{vandenHorn&vanWeert81,vandenHorn&vanWeert84}). In the limit $\eta_\nu \gg 1$ and at fixed temperature, degeneracy increases the viscosity as $\nu_\nu \propto \eta_\nu^2$ \citep{vandenHorn&vanWeert81}.
In the following we give a brief summary of this result and provide an analytic estimate of the viscosity in the degenerate regime.

Following the semi-classical extension of the Chapman-Enskog procedure to Fermions by \cite{Uehling&Uhlenbeck33},
\cite{vandenHorn&vanWeert84} derive the neutrino viscosity---including the effects of degeneracy---in the diffusive regime, where the optical depth is large.
This viscosity depends on the effective neutrino mean free path
\begin{equation}
    \lambda^*_\nu(\epsilon) \equiv \left[ \kappa^*_{\rm a}(\epsilon) + \kappa_{2}(\epsilon) \right]^{-1}
\end{equation}
as a function of neutrino energy $\epsilon$. 
Here $\kappa^*_{\rm a}$ is the absorption opacity\footnote{In this context, `opacity' is defined as an inverse mean free path.} corrected for stimulated absorption, and $\kappa_{2}$ is the appropriately angle-averaged scattering opacity
defined via
\begin{equation}
    \kappa_n(\epsilon) = \int d\Omega \frac{d \kappa_{\rm s} \left( \epsilon, \cos\theta \right) }{d \Omega}
    \begin{cases}
    1 &, n=0
    \\
    1 - P_n(\cos\theta)  &, n \geq 1
    \end{cases}
    ,
\end{equation}
where $P_n(x)$ is the Legendre polynomial of order $n$. 
Note that the corresponding transport (momentum transfer) opacity $\kappa_1$ is in general not equal to $\kappa_2$.
To a good approximation, neutrino--nucleon scattering obeys $d \kappa_{\rm s}/d\Omega = (\kappa_0 / 4\pi) (1 + \delta \cos\theta)$ \citep{Burrows+2006} so the relevant scattering opacity is $\kappa_2 = \kappa_0$.

The kinematic shear viscosity of neutrinos is then given by \citep{vandenHorn&vanWeert84}
\begin{equation}
\label{eq:Appendix_nunu_general}
\nu_\nu = \sum_{\nu_i \in \left\{\nu_e, ... \right\}}  \frac{4 \pi}{15} \frac{(kT)^4}{h^3 c^4 \rho} \int_0^\infty \lambda^*_{\nu_i}(x kT) \frac{x^4 e^{x-\eta_{\nu_i}}}{(e^{x-\eta_{\nu_i}} + 1)^2} \,dx 
= \sum_{\nu_i \in \left\{\nu_e, ... \right\}}  \frac{4}{15} \frac{e_{\nu_i}}{\rho c} \left\langle \lambda^*_{\nu_i} \right\rangle
,
\end{equation} 
where $\rho$ is the fluid density, $T$ the neutrino temperature, $\eta_\nu$ the neutrino degeneracy parameter, and the sum runs over the six neutrino species $\nu_i = \nu_e, \bar{\nu}_e, \nu_\mu, ...$.
The integrand is proportional to $f_{\rm eq} (1 - f_{\rm eq})$ and therefore accounts for final-state (Pauli) blocking, where $f_{\rm eq} = (e^{\epsilon/kT - \eta_\nu}+1)^{-1}$ is the equilibrium Fermi-Dirac neutrino distribution function.
In the final line, the neutrino viscosity has been rewritten in a more familiar form---as a function of the neutrino energy density $e_\nu$ and an appropriately averaged mean free path,
\begin{equation}
\label{eq:Appendix_average_mfp}
    \left\langle \lambda^*_\nu \right\rangle \left( T,\eta_\nu \right)
    \equiv 
    \frac{\int d\epsilon\, \lambda^*_\nu(\epsilon) \epsilon^4 f_{\rm eq} (1-f_{\rm eq})}{\int d\epsilon\, \epsilon^4 f_{\rm eq} (1-f_{\rm eq})}
    = \frac{\int_0^\infty dx\, \lambda^*_\nu(x kT) x^4 {e^{x-\eta_\nu}}{\left( e^{x-\eta_\nu} +1 \right)^{-2}}}{\int_0^\infty dx\, x^4 {e^{x-\eta_\nu}}{\left( e^{x-\eta_\nu} +1 \right)^{-2}}}
    .
\end{equation}
Note that the denominator above may be integrated by parts to obtain $\int x^4 f_{\rm eq} (1-f_{\rm eq}) dx = 4 \int x^3 f_{\rm eq} dx \propto e_\nu$.
Although the \cite{vandenHorn&vanWeert84} result (\ref{eq:Appendix_nunu_general},\ref{eq:Appendix_average_mfp}) is frequently quoted in the literature---we find that there are prevailing misinterpretations of the appropriately averaged mean free path (including in a previous version of this manuscript). We comment on these below.

\subsection{Comparison with Other Work}

As discussed in \cite{vandenHorn&vanWeert83}, Eq.~(\ref{eq:Appendix_average_mfp}) can be viewed as a Rosseland mean if one identifies an effective neutrino analog to the Planck function as $B_\nu \propto \epsilon^3 f_{\rm eq}$.
Because $\left\langle \lambda^*_\nu \right\rangle$ is often referred to as the `energy-averaged' mean free path (a term first used in the original work of \citealt{vandenHorn&vanWeert84}), it may be confused and incorrectly calculated as $\left\langle \lambda^*_\nu \right\rangle \propto \int \lambda^*_\nu(\epsilon) \epsilon^3 f_{\rm eq} \,d\epsilon$. This mistaken interpretation is off by one factor of $\epsilon$ and neglects the Pauli blocking term $(1-f_{\rm eq})$ that is important when neutrinos are degenerate. In general, neutrino degeneracy---one of the main points in the work of \cite{vandenHorn&vanWeert81,vandenHorn&vanWeert84}---has often been neglected in more recent calculations of neutrino viscosities.

For the purpose of facilitating a more direct comparison with previous work, we consider an effective mean free path of the form $\lambda^*_\nu = \tilde{\lambda}(T) \left(\epsilon / kT\right)^{-2}$, as relevant for neutrino--nucleon scattering.
In the limit of vanishing chemical potential one obtains from Eq.~(\ref{eq:Appendix_average_mfp}) an average mean free path $\left\langle \lambda^*_\nu \right\rangle(T,\eta_\nu=0) = (5/7\pi^2) \tilde{\lambda}(T) \simeq 0.072 \tilde{\lambda}(T)$. The modified expression given by \cite{Guilet+15} for the neutrino viscosity (their Eqs.~8,9) is consistent with this result. However the expression in \cite{Guilet+15} is {\it not} correct for a more general energy dependence of 
$\lambda^*_\nu(\epsilon)$
or when the neutrino chemical potential is non-zero
because their $\left\langle \lambda^*_\nu \right\rangle_{\rm Guilet15} \propto \int \lambda^*_\nu \epsilon^3 f_{\rm eq} \,d\epsilon$.
The expressions in \citet[their Eqs.~5--7]{Thompson+05} imply $\left\langle \lambda^*_\nu \right\rangle_{\rm Thompson05} \propto \int \lambda^*_\nu \epsilon^2 f_{\rm eq} \,d\epsilon$, nominally off by a factor of $\epsilon^{2}$ and seemingly missing the Pauli blocking term $(1-f_{\rm eq})$. However, Pauli blocking is in fact implicitly included in the calculations of \cite{Thompson+05} because their energy-dependent neutrino mean free paths already account for final-state blocking corrections (T.~Thompson 2022, private communication). Furthermore, in their numerical calculations \cite{Thompson+05} calculate $\tilde{J}_\nu$ as the zeroth angular moment of the neutrino {\it specific intensity} 
rather than the distribution function (T.~Thompson 2022, private communication). This implies that $\tilde{J}_\nu \propto \epsilon^3 f_{\rm eq}$, rather than $\tilde{J}_\nu \propto f_{\rm eq}$ as suggested by Eq.~(7) of \cite{Thompson+05}.
Therefore, $\left\langle \lambda^*_\nu \right\rangle_{\rm Thompson05} \propto \int \lambda^*_\nu \epsilon^5 f_{\rm eq} (1-f_{\rm eq}) \,d\epsilon$ and is only off by one factor of~$\epsilon$. For the $\lambda_\nu^* \propto \epsilon^{-2}$ case considered above and at zero chemical potential, this implies $\left\langle \lambda^*_\nu \right\rangle_{\rm Thompson05} = (147 / 310\pi^2) \tilde{\lambda}(T) \simeq 0.048 \tilde{\lambda}(T)$, a modest $\sim 30\%$ smaller than the \cite{vandenHorn&vanWeert84} result (Eq.~\ref{eq:Appendix_average_mfp}).

\subsection{Analytic Estimate of Viscosity at Arbitrary Neutrino Degeneracy}

Using analytic approximations to the neutrino opacity we now obtain a numerical estimate of $\nu_\nu$ defined in Eq.~(\ref{eq:Appendix_nunu_general}).
Adopting cross-sections from \cite{Burrows+2006}, ignoring absorption, and assuming that $\nu-n$ and $\nu-p$ scattering dominate the scattering opacity (with an $\epsilon^2$ dependence of the cross-sections), we find that
\begin{equation}
\nu_\nu = 
\sum_{\nu_i \in \left\{\nu_e, ... \right\}}  \frac{2 m_e^2 m_p}{15 \pi^2 \hbar^3 \sigma_0} \left(\frac{kT}{\rho}\right)^2 f(Y_{e})
\int_0^\infty x^2 \frac{e^{x-\eta_{\nu_i}}}{(e^{x-\eta_{\nu_i}} + 1)^2} \,dx
,
\end{equation}
where $\sigma_0 \simeq 1.705 \times 10^{-44} \, {\rm cm}^2$ \citep{Burrows+2006} and 
\begin{equation}
\label{eq:Appendix_fYe}
    f(Y_{e}) \equiv 
    \left[ \frac{1+3 g_{A}^2}{4} + \left( 4 \sin^4 \theta_{W} - 2 \sin^2 \theta_{W} \right)
    Y_{e} \right]^{-1}
    \simeq \left( 1.471 - 0.248 Y_{e} \right)^{-1}
\end{equation}
is a (weak) function of the electron fraction $Y_{e}$ that encapsulates the relative contribution of $\nu-n$ and $\nu-p$ scattering to the total cross section. 
Here $\theta_W$ is the Weinberg angle, $\sin^2 \theta_W \simeq 0.231$, and $g_A \simeq -1.276$ is the axial-vector coupling constant \citep{Markisch+19,ParticleDataGroup20}.
Assuming all neutrinos have the same temperature and that beta equilibrium enforces $\eta_{\bar{\nu}_i} = - \eta_{\nu_i}$, we can write a closed-form analytic solution to the integral
\begin{equation}
  \int_0^\infty x^2 \left[ \frac{e^{x-\eta_{\nu_i}}}{(e^{x-\eta_{\nu_i} } + 1)^2} +
  \frac{e^{x - \eta_{\bar{\nu}_i}}}{(e^{x-\eta_{\bar{\nu}_i}} + 1)^2}
  \right] \,dx
  = \frac{\pi^2}{3} + \eta_{\nu_i}^2 .
\end{equation}   
Using this result, we obtain a final expression for the kinematic shear viscosity of neutrinos in the diffusive regime,
\begin{eqnarray}
\label{eq:Appendix_nunu_final}
  \nu_\nu 
  &=& \frac{2 m_e^2 m_p}{15 \hbar^3 \sigma_0} \left(\frac{kT}{\rho}\right)^2 f(Y_e) 
  \left[ 1 + 
  \frac{\eta_{\nu_e}^2 + \eta_{\nu_\mu}^2 + \eta_{\nu_\tau}^2}{\pi^2} 
  \right]
  \nonumber\\
  &\approx& 1.66 \times 10^{10} \, {\rm cm}^2 \, {\rm s}^{-1} \, \left(\frac{T}{10 \, \textrm{MeV}}\right)^2 \left(\frac{\rho}{10^{13} \, {\rm g \, cm}^{-3}}\right)^{-2}
  \frac{f(Y_e)}{0.7}
  \left[ 1 + 
  \frac{\eta_{\nu_e}^2 + \eta_{\nu_\mu}^2 + \eta_{\nu_\tau}^2}{\pi^2} 
  \right]
  .
\end{eqnarray}
When neutrinos are non-degenerate ($\eta_\nu \approx 0$), this result is in broad agreement with the \cite{Keil+96} estimate that is commonly used in the literature, albeit $\sim$40\% larger (\citealt{Keil+96} find $\nu_\nu \approx 1.2 \times 10^{10} \, {\rm cm}^2\,{\rm s}^{-1}$ for the same temperature and density).
The importance of Eq.~(\ref{eq:Appendix_nunu_final}) is its applicability to regions with arbitrary neutrino degeneracy.
As shown by \cite{vandenHorn&vanWeert81,vandenHorn&vanWeert84}, neutrino degeneracy increases the viscosity $\propto \eta_\nu^2$.

In the core of BNS merger remnants or PNS, electron neutrino degeneracy can become appreciable.
A rough estimate of the $\nu_e$ degeneracy parameter can be obtained by equating the neutrino energy density 
$e_{\nu_e} = 4\pi (hc)^{-3} (kT)^4 F_3(\eta_{\nu_e})$
to the trapped thermal energy $E$. 
Here $F_n(\eta) \equiv \int_0^\infty x^n \left(e^{x-\eta}+1\right)^{-1} dx$ is the Fermi integral, which satisfies $F_n(\eta_\nu \gg 1) \approx \eta_\nu^{n+1}/(n+1)$.
In the limit $\eta_{\nu_e} \gg 1$, this gives $\eta_{\nu_e} \sim 18\, (T/10\,{\rm MeV})^{-1} (E/10^{53}\,{\rm erg})^{1/4} (R/20\,{\rm km})^{-3/4}$.
A more accurate estimate may be given in terms of local variables. Using the electron neutrino number density 
$n_{\nu_e} = 4\pi (hc)^{-3} (kT)^3 F_2(\eta_{\nu_e})$
in the limit $\eta_{\nu_e} \gg 1$ we find that
\begin{equation}
\label{eq:Appendix_eta}
    \eta_{\nu_e} 
    \underset{\eta_{\nu_e} \gg 1}{\approx}
    \left( \frac{3 n_{\nu_e}}{4\pi} \right)^{1/3} \frac{hc}{kT}
    \approx 14 \, \left(\frac{T}{10\,{\rm MeV}}\right)^{-1} \left(\frac{\rho}{10^{14}\,{\rm g\,cm}^{-3}}\right)^{1/3} \left(\frac{Y_{\nu_e}}{0.1}\right)^{1/3} ,
\end{equation}
where $Y_{\nu_e}$ is the electron neutrino number fraction.
This expression is in good agreement with the neutrino degeneracy parameter that is found in the PNS evolution models we have studied.
Eq.~(\ref{eq:Appendix_eta}) implies a factor $\sim\mathcal{O}(10)-\mathcal{O}(100)$ enhancement to the neutrino viscosity deep within the neutron star core.
Specifically, using Eqs.~(\ref{eq:Appendix_nunu_final},\ref{eq:Appendix_eta}) we find in the deeply degenerate regime
\begin{equation}
\label{eq:nunu_degenerate}
    \nu_\nu \underset{\eta_{\nu_e} \gg 1}{\approx}
    3.28 \times 10^{9} \,{\rm cm^{2}\,s^{-1}}\,
    \left(\frac{\rho}{10^{14}\,{\rm g\,cm}^{-3}}\right)^{-4/3}
    \left(\frac{Y_\nu}{0.1}\right)^{2/3} \frac{f(Y_e)}{0.7}
    .
\end{equation}

\bibliographystyle{aasjournal}
\bibliography{refs}


\end{document}